\documentclass[fleqn]{mnras}

\usepackage{amsmath}
\let\bibhang\relax
\usepackage{natbib}
\usepackage{gensymb}
\usepackage{graphicx}
\usepackage[utf8x]{inputenc}
\usepackage{appendix}

\usepackage{rotating} 
\usepackage[font=normalsize, labelfont=bf]{caption}

\graphicspath{{./}{figures/}}

\title[Diffuse X-ray luminosity and radio power in groups]{The relation between the diffuse X-ray luminosity and the radio power of the central AGN in galaxy groups}

\author[T. Pasini, M. Brüggen, F. de Gasperin et al.]{ T. Pasini$^{1}$\thanks{E-mail: thomas.pasini@hs.uni-hamburg.de},
M. Brüggen$^{1}$,
F. de Gasperin$^{1}$,
L. Bîrzan,$^{1}$,
E. O'Sullivan$^{2}$,
\newauthor
A. Finoguenov$^{3}$,
M. Jarvis$^{4,5}$,
M. Gitti$^{6,7}$,
F. Brighenti$^{6}$,
I. H. Whittam$^{4,5}$,
\newauthor
J. D. Collier$^{8,9}$,
I. Heywood$^{5,10,11}$
and G. Gozaliasl$^{3}$
\\
$^{1}$Hamburger Sternwarte, Universität Hamburg, Gojenbergsweg 112, 21029 Hamburg, Germany\\
$^{2}$Harvard-Smithsonian Center for Astrophysics, 60 Garden Street, Cambridge, MA02138, USA\\
$^{3}$Department of Physics, University of Helsinki, P. O. Box 64, FI-00014 , Helsinki, Finland\\
$^{4}$Department of Physics and Astronomy, University of the Western Cape, Robert Sobukwe Road, Bellville 7535, South Africa\\
$^{5}$Astrophysics, University of Oxford, Denys Wilkinson Building, Keble Road, Oxford, OX1 3RH\\
$^{6}$Dipartimento di Fisica e Astronomia (DIFA), Universita‘ di Bologna, via Gobetti 93/2, I-40129 Bologna, Italy\\
$^{7}$Istituto Nazionale di Astrofisica (INAF)—Istituto di Radioastronomia (IRA), via Gobetti 101, I-40129 Bologna, Italy\\
$^{8}$The Inter-University Institute for Data Intensive Astronomy (IDIA), Department of Astronomy, University of Cape Town, \\ Private Bag X3, Rondebosch, 7701, South Africa\\
$^{9}$School of Science, Western Sydney University, Locked Bag 1797, Penrith, NSW 2751, Australia\\
$^{10}$Department of Physics and Electronics, Rhodes University, PO Box 94, Makhanda, 6140, South Africa\\
$^{11}$South African Radio Astronomy Observatory, 2 Fir Street, Black River Park, Observatory, Cape Town 7925, South Africa\\
}

\date{Accepted 2020 July 8. Received 2020 July 8; in original form 2020 May 6}

\begin{document}
\label{firstpage}
\pagerange{\pageref{firstpage}--\pageref{lastpage}}
\maketitle

\begin{abstract}

Our understanding of how AGN feedback operates in galaxy clusters has improved in recent years owing to large efforts in multi-wavelength observations and hydrodynamical simulations. However, it is much less clear how feedback operates in galaxy groups, which have shallower gravitational potentials.
In this work, using very deep VLA and new MeerKAT observations from the MIGHTEE survey, we compiled a sample of 247 X-ray selected galaxy groups detected in the COSMOS field. We have studied the relation between the X-ray emission of the intra-group medium and the 1.4 GHz radio emission of the central radio galaxy. For comparison, we have also built a control sample of 142 galaxy clusters using $ROSAT$ and NVSS data. 
We find that clusters and groups follow the same correlation between X-ray and radio emission. Large radio galaxies hosted in the centres of groups and merging clusters increase the scatter of the distribution. Using statistical tests and Monte-Carlo simulations, we show that the correlation is not dominated by biases or selection effects. We also find that galaxy groups are more likely than clusters to host large radio galaxies, perhaps owing to the lower ambient gas density or a more efficient accretion mode. In these groups, radiative cooling of the ICM could be less suppressed by AGN heating.
We conclude that the feedback processes that operate in galaxy clusters are also effective in groups.
\end{abstract}


\section{Introduction} 
\label{sec:intro}

Over the past years, deep observations performed by both \textit{Chandra} and \textit{XMM-Newton} have led the way for a better understanding of the X-ray emission produced by the Intra-Cluster Medium (ICM), its consequent cooling and how this links to structure formation. Particularly, multi-wavelength studies based on the combination of X-ray and radio observations have shown that the ICM and the Active Galactic Nucleus (AGN) usually hosted in the Brightest Cluster Galaxy (BCG) are part of a tight cycle in which the cooling of the hot ($\sim$ 10$^7$ K) ICM is regulated by the mechanical feedback provided by the AGN itself (see e.g. reviews by \citealt{McNamara-Nulsen_2012, Gitti_2012}). In this scenario, AGN jets and outflows produce shock waves and cold fronts \citep{McNamara_2000, Fabian_2006} and inflate bubbles in the ICM, known as X-ray cavities, that can be used to assess the AGN mechanical power \citep{Birzan_2004, Rafferty_2006}, establishing a tight feedback cycle with the diffuse gas.
The impact of the AGN could be even stronger in galaxy groups, where the gravitational potential is shallower. Here, even a relatively small energy injection could eject gas from the group itself \citep{Giodini_2010}. It has been suggested that AGN feedback could thus set apart galaxy clusters from groups, especially in terms of their baryonic properties \citep{Jetha_2007}, albeit there is still no universal agreement in the distinction between these objects. 

Therefore galaxy groups, that are the repositories of the majority of baryons and host more than half of all galaxies \citep{Eke_2006}, are key in order to reach a complete understanding of the AGN feedback cycle and of how it is able to influence the evolution of galaxies and their environments \citep[e.g.,][]{Giacintucci_2011b}. 
However, X-ray observations of galaxy groups are relatively difficult since most of them lie at the lower sensitivity limit of the current generation of instruments. This normally prevents us from reaching the combination of signal-to-noise ratio and resolution required to perform the same type of analysis that is usually applied to galaxy clusters (see e.g., \citealt{Willis_2005}). 
Nonetheless, there are numerous studies of galaxy groups that make use of either deep observations, and/or low-redshift samples. \citet{O'Sullivan_2017} presented an optically-selected, statistically complete sample of 53 low-redshift galaxy groups (Complete Local Volume Groups Sample, CLoGS), for which they were able to perform a detailed X-ray analysis. They classified groups into cool-cores and non cool-cores and studied the central radio galaxy \citep{Kolokythas_2018}, finding that $\sim$ 92\% of their groups' dominant galaxies host radio sources. Other studies are based on estimating the scaling relations between observables such as temperature, luminosity, entropy and mass \citep[e.g.,][]{Lovisari_2015}. \citet{Bharadwaj_2014} estimated the central cooling time (CCT; see Sec. \ref{sec:cooltime} for a definition) in a sample of galaxy groups and found that the fractions of strong (CCT $<$ 1 Gyr), weak (1 $<$ CCT $<$ 7.7 Gyr), and non cool-cores (CCT $>$ 7.7 Gyr) were similar to those in galaxy clusters. They also found that BGGs (Brightest Group Galaxies) in their galaxy groups may have a higher stellar mass than BCGs in clusters.

Simulations by \citet{Gaspari_2011b} showed that AGN feedback may be more persistent and delicate in galaxy groups than in galaxy clusters. A small number of deep observations of single, local objects also detected cavities and shocks \citep[e.g.,][]{Nulsen_2005c, Gitti_2010, Randall_2015, Forman_2017}, allowing investigations of the balance between the AGN energy injection and the gas cooling. X-ray cavities were also recently observed by \citet{Birzan_2020} in a sample of 42 systems, of which 17 are groups or ellipticals.
\\ \indent
\citet{Ineson_2013, Ineson_2015} performed a study of the interactions between AGN and their environment in a sample of radio-loud AGN in clusters and groups. They found a correlation between the X-ray emission from the intra-group medium and the 151 MHz power of the central radio source. They also argued that such a correlation could arise from AGN in a phase of radiatively-inefficient accretion (Low Excitation Radio Galaxies, or LERGs), while High Excitation Radio Galaxies (HERGs) stand out of the distribution and show higher radio powers. The origin of such a relation is not obvious. In fact, X-ray emission in clusters and groups is mostly due to line emission and Bremsstrahlung, that consequently allow the ICM to cool from high ($\sim$ 10$^7$ K) temperatures. The time scale of such radiative losses is thus strongly dependent on the distance of the diffuse gas from the cluster (or group) core, varying from less than 1 Gyr in the centre of the strongest cool-cores to $\sim$ few Gyrs moving towards the outskirts.
On the other hand, \citet{Nipoti-Binney_2005} suggested that the AGN power output could act in cycles of $\sim$ 10$^8$ yr. Hence, the time scales of these two processes are usually significantly different.
However, \citet{O'Sullivan_2017} found that groups typically show shorter cooling time at a given radius when compared to clusters, due to the high cooling efficiency of line emission at kT $<$ 2 keV. 

Here, we study the relationship between the X-ray emission from the intra-group medium and the radio emission from the radio galaxy hosted in the centre of a large (N = 247) sample of X-ray detected galaxy groups in the 2 deg$^2$ of the COSMOS field (RA = 10$^\text{h}$00$^\text{m}$28.6$^\text{s}$, DEC = +02$\degree$12$^\text{m}$21.0$^\text{s}$, J2000). This field was chosen as it offers a unique combination of deep and multi-wavelength data. In order to account for the faintness of groups, we make use of the deepest observations and catalogs.

This paper is organized as follows: in Sec.~\ref{sec:gensample}, we explain how we built the catalog and we describe its main properties. In Sec.~\ref{sec:analysis}, we compare the groups with a sample of galaxy clusters and explore the correlation between the X-ray and the radio emission exploiting statistical tests and a Monte-Carlo simulation. In Sec.~\ref{sec:discussion}, we discuss the physical implications of our results and put them into context. In Sec.~\ref{sec:conclusions}, we draw the conclusions. Throughout this paper, we adopt a fiducial $\Lambda$CDM cosmological model with $H_0$ = 71 km s$^{-1}$ Mpc$^{-1}$, $\Omega_\Lambda$ = 0.7 and $\Omega_M $ = 0.3.

%%%%%%%%%%%%%%%%%%%%%%%%%%%%%%%%%%%%%%%%%%%%%%%%%%%%%%%%%%%%%%%%%%%%%%%%%%%%%%%%%%%%%%%%%%%%

\section{The samples}
\label{sec:gensample}

\subsection{Construction of the group sample}
\label{sec:sample}

\citet{Gozaliasl_2019} presented a catalog of 247 X-ray-selected galaxy groups in the COSMOS field, obtained combining all available \textit{Chandra} and \textit{XMM-Newton} observations, in the redshift range (spectroscopic for 183 groups, photometric for the remaining 64) $0.08 \leq  z  \leq 1.75$ and with luminosities in the $0.1-2.4$~keV band ranging from $\sim$ 10$^{41}$ to $\sim$ 10$^{44}$ erg s$^{-1}$. The flux limit of the sample is $\sim$ 3 $\cdot 10^{-16}$ erg s$^{-1}$ cm$^{-2}$. Setting a search radius of 30$\arcsec$ from the groups' centre (assumed as coincident with the X-ray peak) and a redshift threshold of $\Delta z$ = 0.02, we cross-matched this sample with the VLA-COSMOS Deep Survey at 1.4 GHz (rms $\sim$ 12 $\mu$Jy beam$^{-1}$, beam = 2.5\arcsec x 2.5\arcsec, \citealt{Schinnerer_2010}), to look for the radio galaxy hosted in the centre of every group. We chose to use VLA-COSMOS to get the highest resolution available. At the mean redshift of our sample ($z \sim$ 0.7), the largest visible angular scale at this frequency corresponds to $\sim$ 450 kpc. The results were then inspected visually, exploiting COSMOS optical catalogs, to check the bounty of the cross-match. We found that, out of the 136 groups for which BGG properties are available, only in 41 ($\sim$ 30\%) the detected radio source is hosted in the BGG. This is consistent with recent works that observed offsets between the optical dominant galaxy and the X-ray peak. \citet{Gozaliasl_2019} found that only 30\% of BGGs in COSMOS are closer to the X-ray peak than 0.1R$_{200}$, and that the peak is often not located at the bottom of the potential well, where the dominant galaxy usually lie. We will return on this in Sec. \ref{sec:clsample}.
%We then inspected the results visually and removed false positives, mostly produced by some groups having more than one central, dominant galaxy. In these cases, we assumed that the BGG is the galaxy hosting the most powerful radio source.

Groups showing no central radio emission were then further inspected exploiting new, deep MeerKAT observations of COSMOS that are part of the MIGHTEE survey (The MeerKAT International GHz Tiered Extragalactic Exploration, \citealt{Jarvis_2016}, {\color{blue} Heywood et al. in prep.}). For these MIGHTEE COSMOS Early Science images, the thermal noise component, measured from the circular polarization images (Stokes V), is 2.2 $\mathrm{\mu}$Jy/beam in the image with 9\arcsec x 7.4\arcsec resolution, and 7.5 $\mathrm{\mu}$Jy/beam in the 4.7\arcsec x 4.2\arcsec resolution image. However, in the map centre the "noise" will appear to be higher due to the contribution from confusion (faint background sources below the formal noise limit). We obtain an estimate of this thermal-plus-confusion noise by subtracting the model of the sky obtained by the PyBDSF source finder \citep{Mohan_2015}\footnote{https://github.com/darafferty/PyBDSF} and generating a RMS map of the image from this residual product using the same software. The mean RMS value over the inner 20\arcsec x 20\arcsec region (where the primary beam attenuation is approximately negligible) is then measured. For the MIGHTEE COSMOS Early Science map this measurement is 4.1 $\mathrm{\mu}$Jy/beam in the 9\arcsec x 7.4\arcsec image, and 8.6 $\mathrm{\mu}$Jy/beam in the 4.7\arcsec x 4.2\arcsec resolution image. The properties of all the radio images are summarized in Table \ref{tab:images}.

\begin{table}%[!htb]
	%\small
	\centering 
	\begin{tabular}{c c c}
		\hline
		\hline
		Image & Beam & Sensitivity \\
		& [arcsec] & [$\mathrm{\mu}$Jy/beam] \\
		\hline
		VLA-COSMOS Deep & 2.5 x 2.5 & 12 \\
		MIGHTEE high-resolution & 4.7 x 4.2 & 8.6 \\
		MIGHTEE low-resolution & 9 x 7.4 & 4.1 \\
		NVSS & 45 x 45 & 450 \\
        \hline
	\end{tabular}
	\caption{Properties of the 1.4 GHz images exploited for the detection of central radio sources in the group sample (see Sec. \ref{sec:sample}) and in the cluster sample (see Sec. \ref{sec:clsample}).}
	\label{tab:images}
\end{table}

Our final sample (hereafter referred to as group sample) consists of 174 (155 detected by both VLA and MIGHTEE and 19 only by MIGHTEE) objects for which we have redshifts and both the X-ray emission (flux, luminosity, R$_{200}$) and the radio emission (flux density, luminosity, largest linear size) from the central radio galaxy, plus 73 groups for which the central source was undetected in the radio band. These are thus treated as $3\sigma$ upper limits, with $\sigma$ being the rms noise of the MIGHTEE low-resolution observation. The largest linear size is defined as the linear size of the major axis of a source, and we will hereafter refer to it as LLS. The 1.4 GHz luminosity was estimated as:

\begin{equation}
\hspace{2.5cm}
    L_{1.4\text{GHz}} = S_{\text{1.4}} 4\pi D_{\text{L}}^2  (1+z)^{-\alpha}
\end{equation}

where S$_{\text{1.4}}$ is the flux at 1.4 GHz, D$_{\text{L}}$ is the luminosity distance at redshift $z$ and $\alpha$ is the spectral index, that was assumed $\sim$ 0.6, since this is the mean synchrotron index usually observed in radio galaxies.

The group sample was further divided into groups with \textit{resolved} and \textit{unresolved} radio galaxies, following the same criteria presented in \citet{Schinnerer_2007} and \citet{Schinnerer_2010}. This classification is based on the assumption that the ratio between integrated and peak flux density gives a measure of the spatial extent of a source in comparison to the size of the synthesized beam(see Sec. 6.2 and appendix of \citealt{Schinnerer_2010} for more details). We find that $\sim$ 32\% (55 objects) of the sample show well-resolved radio sources, while for the remaining $\sim$ 68\% (119 objects) we only have upper limits on the LLS. Exploiting MIGHTEE images, we found that no sources are spatially unextended due to VLA lacking surface brightness sensitivity. Table \ref{tab:sample} summarizes the composition of the sample; all the properties of the objects are listed in Table \ref{tab:complete}, described in Appendix \ref{appendix} and available as online material.

\begin{table}%[!htb]
	%\small
	\centering 
	\begin{tabular}{c c c c}
		\hline
		\hline
		& Total & VLA + MIGHTEE & MIGHTEE \\
		\hline
        Detections & 174 & 155 & 19 \\
        Upper limits & 73 & & $>$ 3$\sigma$ \\
        Resolved & 55 & 50 & 5 \\
        Unresolved & 119 & 119 & 14 \\
        \hline
	\end{tabular}
	\caption{Composition of the group sample. In a sample of 247 galaxy groups, we observed a central radio source in 174 of them. 155 are detected by both VLA-COSMOS and MIGHTEE, while 19 only by MIGHTEE. Among the 155 VLA+MIGHTEE detections, 55 are resolved, while 119 remain unresolved. Among the 19 MIGHTEE-only detections, 5 are resolved and 14 remain unresolved.}
	\label{tab:sample}
\end{table}

\subsection{Characteristics of the group sample}

\begin{figure}
	%\hspace*{-6.9cm} 
	\centering
	\includegraphics[height=27em, width=28em]{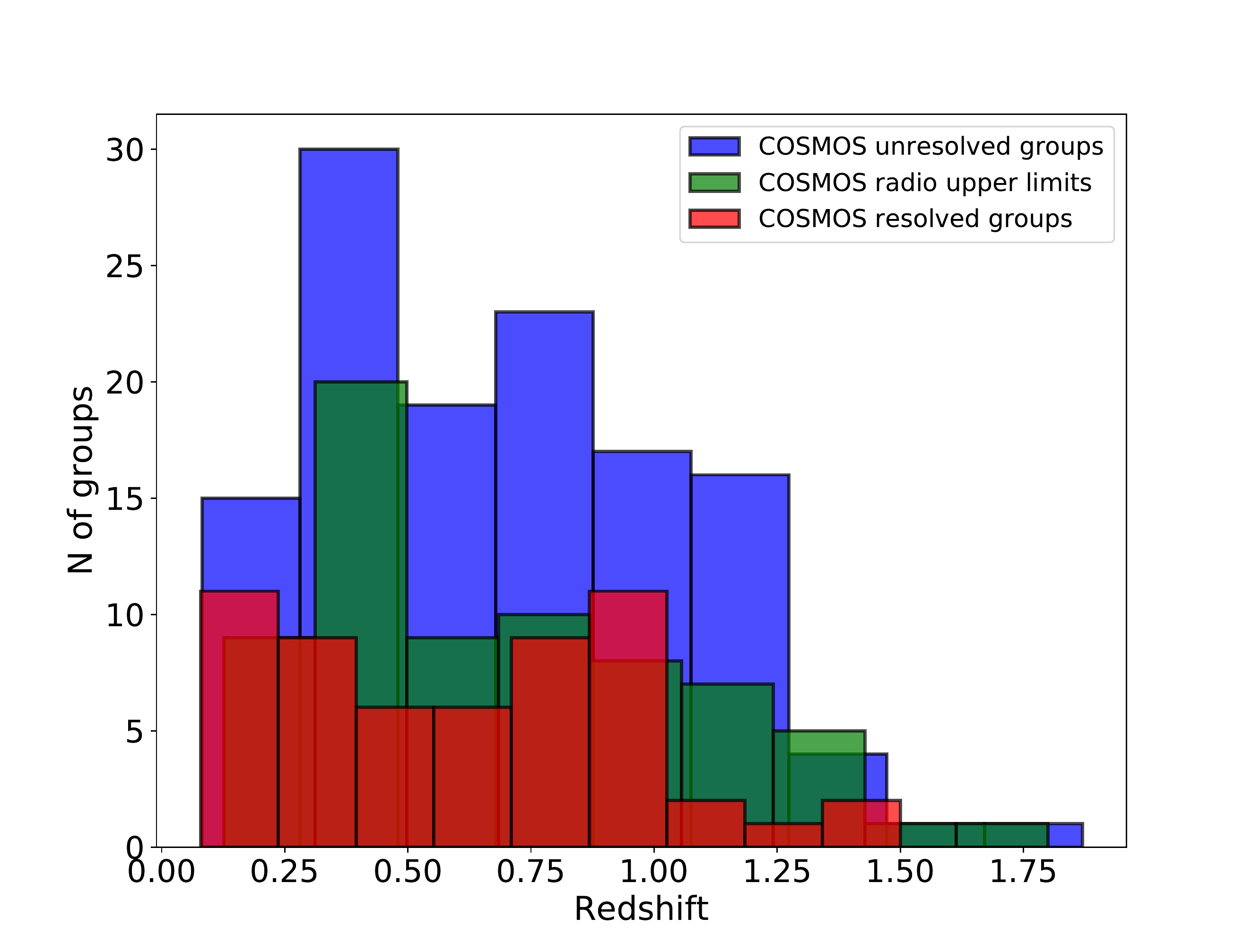}
	%\hspace*{0.2cm}
	\includegraphics[height=27em, width=28em]{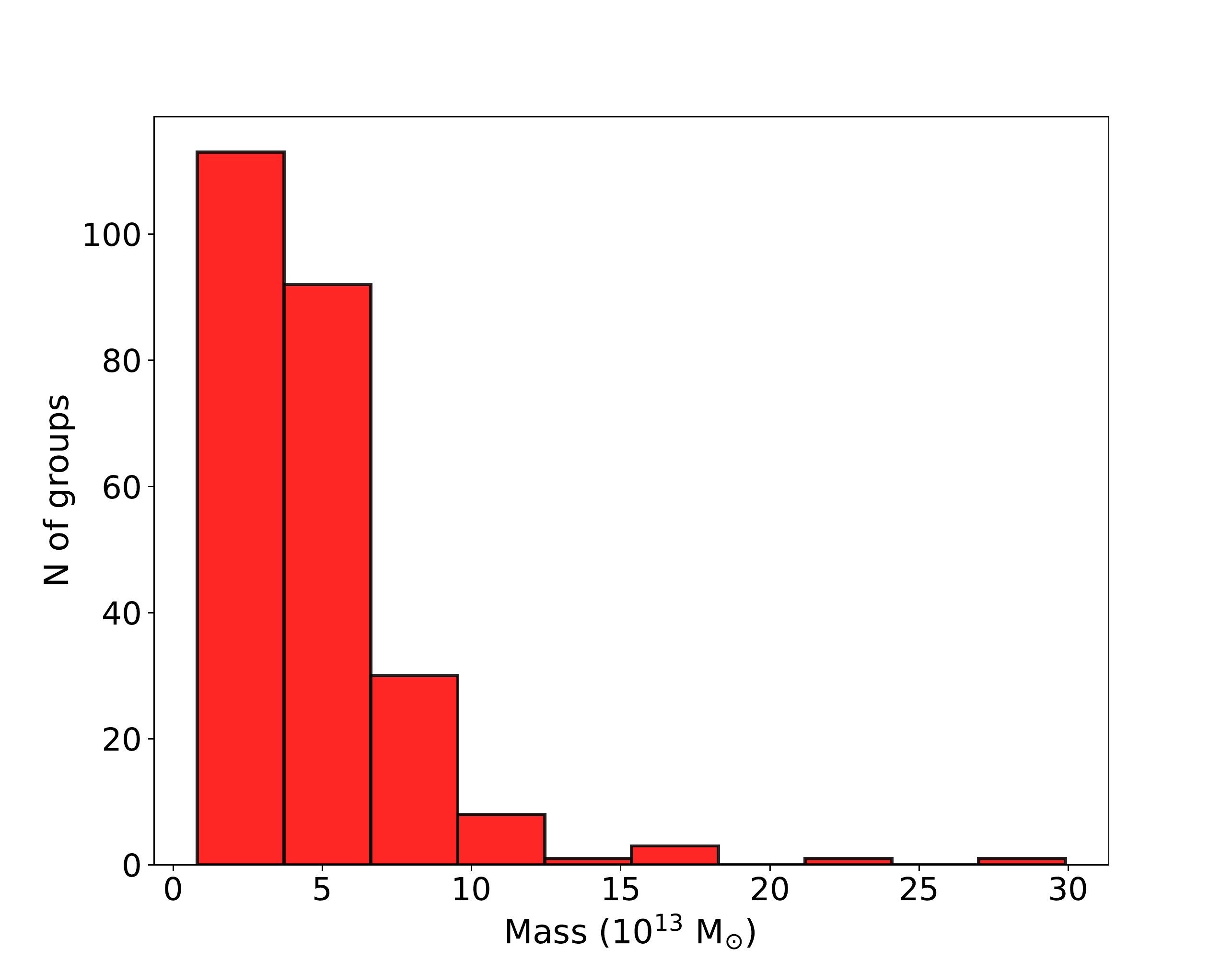}
	\caption{\textit{Top Panel}: Histogram showing the redshift distribution of the group sample, classified into objects with resolved (red) and unresolved (blue) radio galaxies and radio upper limits (green). \textit{Bottom Panel}: Mass (M$_{200}$) distribution of the group sample. The mass was estimated through the L$_{\text{X}}$ - M$_{200}$ correlation by \citet{Leauthaud_2010}.}
	\label{fig:reddistr}
\end{figure}

Fig. \ref{fig:reddistr} shows the redshift and mass (M$_{200}$, the mass within the radius corresponding to 200 times the critical density) distributions of the group sample. The redshift distribution of resolved and unresolved radio sources show a similar behaviour: we detect both resolved, high-redshift and unresolved, low-redshift radio galaxies. The masses of most groups, estimated via the L$_{\text{X}}$ - M$_{200}$ correlation \citep{Leauthaud_2010}, lie within $\sim$ 10$^{13}$ and $\sim$ 10$^{14}$ M$_\odot$.
%, showing that this classification is not entirely produced by unresolved objects being more distant, but also by an intrinsic difference in the radio galaxy linear size.

\begin{figure}
	%\hspace*{-6.9cm} 
	\centering
	\includegraphics[height=27em, width=28em]{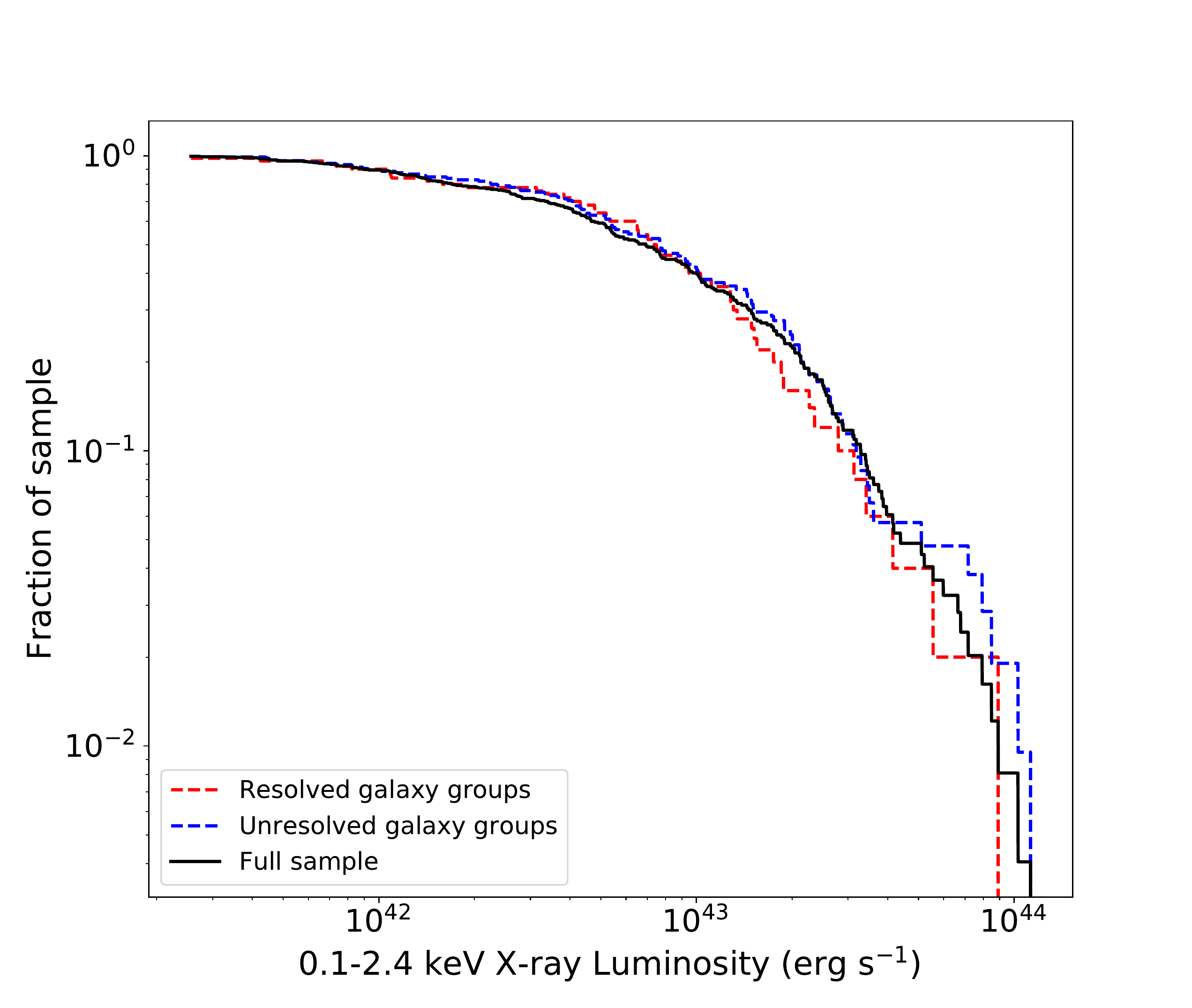}
	%\hspace*{0.2cm}
	\includegraphics[height=27em, width=28em]{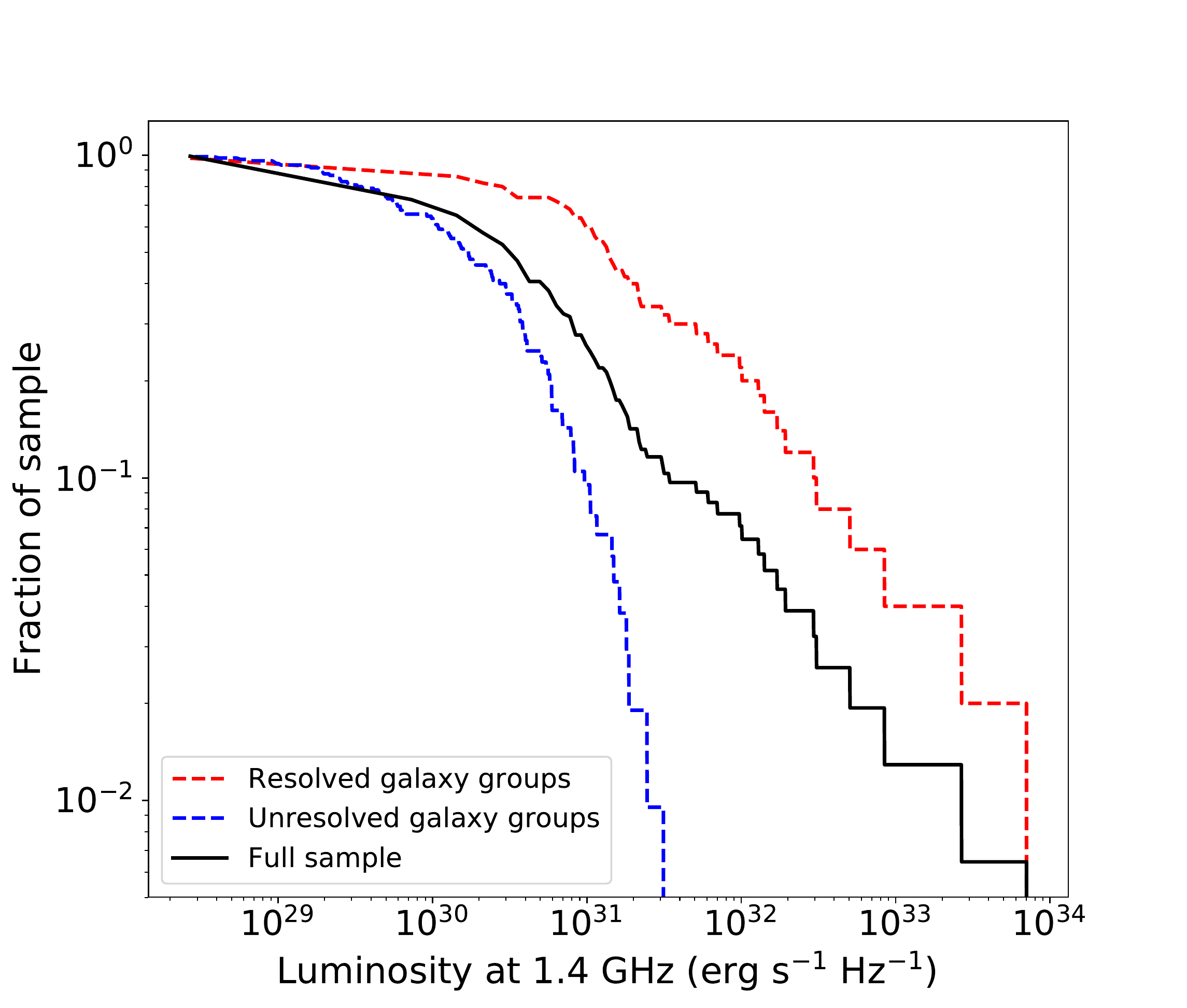}
	\caption{\textit{Top Panel}: X-ray luminosity distribution function for the group sample in the 0.1-2.4 keV band. The red line denotes groups hosting resolved VLA radio sources, the blue line unresolved ones, while the black line represents the full sample. \textit{Bottom Panel}: 1.4 GHz luminosity distribution function for the group sample. The red line denotes groups hosting resolved VLA radio sources, the blue line unresolved ones, while the black line represents the full sample.}
	\label{fig:lumfunct}
\end{figure}

In Fig.~\ref{fig:lumfunct} we show the X-ray and radio luminosity distribution functions of the group sample.
The X-ray distribution function suggests that there is no significant difference in the intra-group medium emission between groups with resolved and unresolved radio galaxies. However, the 1.4 GHz function shows that resolved radio sources are able to reach higher powers at these frequencies, while unresolved radio galaxies exhibit a break between 10$^{31}$ and 10$^{32}$ erg s$^{-1}$ Hz$^{-1}$. We also note that the radio distribution function of our groups' sources spans the same luminosity range as BCGs in clusters \citep[e.g.,][]{Hogan_2015, Yuan_2016}, despite the different environments. 

\begin{figure}
	%\hspace*{-6.9cm} 
	\centering
	\includegraphics[height=29em, width=29em]{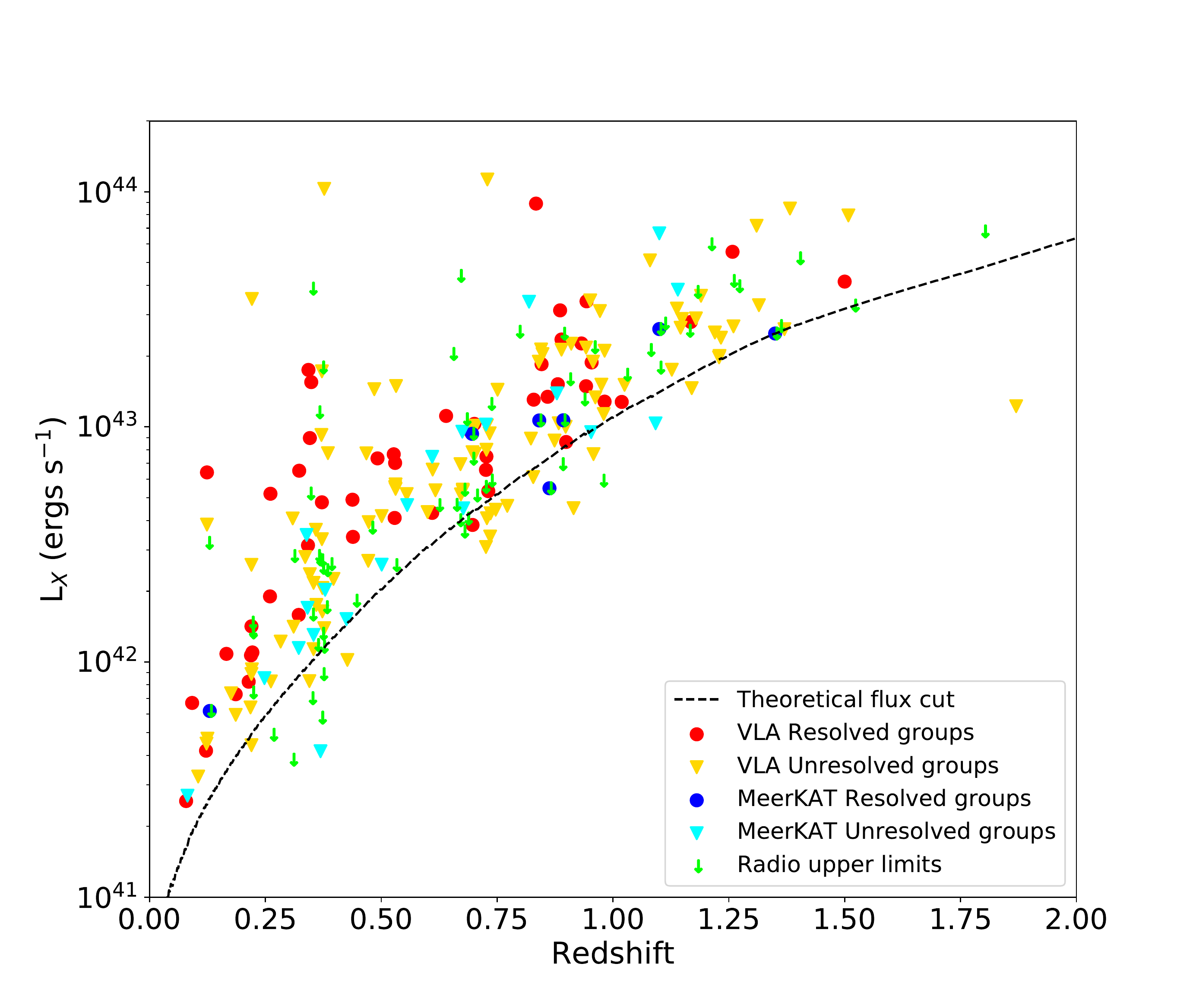}
	\caption{Malmquist bias for the group sample: X-ray luminosity in the 0.1-2.4 keV band vs. redshift. Circles and triangles represent groups hosting resolved and unresolved radio sources, respectively. Green arrows denote upper limits in the radio band. The dashed line represents the theoretical cutoff of the flux-limited sample, produced assuming a surface brightness of $\sim$ 10$^{-15}$ erg s$^{-1}$ cm$^{-2}$ arcmin$^{-2}$ in the 0.5-2 keV band in a circle with a radius of 32\arcsec, which corresponds to the construction of the COSMOS catalog.
	Some objects lie under the cutoff, since in some of the COSMOS regions the X-ray sensitivity was higher due to different sky coverages.}
	\label{fig:malquist}
\end{figure}

Since our sample is flux-limited, it shows the typical Malmquist bias displayed in Fig.~\ref{fig:malquist}. At low redshifts we have mostly low X-ray luminosity objects, while groups with high luminosities are rarer. As we move to higher redshifts, we start to see more powerful groups, while weak objects disappear due to their faintness. The fact that the distribution is not perfectly aligned with the curve is likely due to how the X-ray catalog was produced. Combining multiple observations, performed with different instruments and different sky coverage, can lead to different flux sensitivities. We will return to this issue in Sec.~\ref{sec:stattests}.

\subsection{A control sample of galaxy clusters}
\label{sec:clsample}

We constructed a sample of galaxy clusters in order to compare it to our main sample of groups. The $ROSAT$ Brightest Cluster Sample (BCS, \citealt{Ebeling_1998}) is a 90\% flux-complete sample of 201 galaxy clusters in the Northern hemisphere with $z\leq 0.3$, that reaches a flux limit of 4.4 $\cdot 10^{-12}$ erg s$^{-1}$ cm$^{-2}$ in the 0.1-2.4 keV band. \citet{Crawford_1999} presented the optical BCG position for 165 of the BCS clusters. We cross-matched it with the NRAO VLA Sky Survey (NVSS, \citealt{Condon_1998}) in order to find the corresponding radio galaxy, using the same criteria discussed in Sec.~\ref{sec:sample}. NVSS has a lower resolution when compared to both VLA-COSMOS and MIGHTEE, but it still provides good estimates of the properties of the radio sources we are interested in (e.g., flux density). The survey is produced with VLA in D configuration, resulting in a largest visible angular scale at 1.4 GHz of 970\arcsec that, at the mean redshift of the CS, corresponds to $\sim$ 2500 kpc. The final catalog (hereafter referred to as cluster sample) consists of 84 galaxy clusters for which we have the same measurements (ICM X-ray luminosity, radio power of the central source, redshift, LLS) as for the group sample, and 58 objects for which we only have upper limits for the radio emission from the central galaxy.

We then exploited the red sequence members of the CODEX cluster sample \citep{Finoguenov_2020} to calculate the fraction of radio galaxies that are BCGs. In clusters, 85 $\pm$ 6\% of the central radio galaxies coincide with the BCG. This is different in galaxy groups, where only $\sim$ 30\% of the central radio sources are hosted by the BGG. In most objects, the X-ray peak should represent a good indicator of the center. \citet{Smolcic_2011} found that, in their sample of groups, radio galaxies are in fact always found next to the X-ray peak, justifying our approach. However, \citet{Gozaliasl_2019, Gozaliasl_2020} showed that BGGs defined within R$_{200}$ have a broad distribution of their group-centric radii, and also show kinematics not well matched to that of simulated central galaxies. It might be that radio sources provide a better identification for central galaxies. This possibility needs to be investigated through multi-wavelength observations and by performing a thorough study of the dynamical properties of central radio galaxies, and will be the subject of a future work.

\begin{figure}
	%\hspace*{-6.9cm} 
	\includegraphics[height=27em, width=28em]{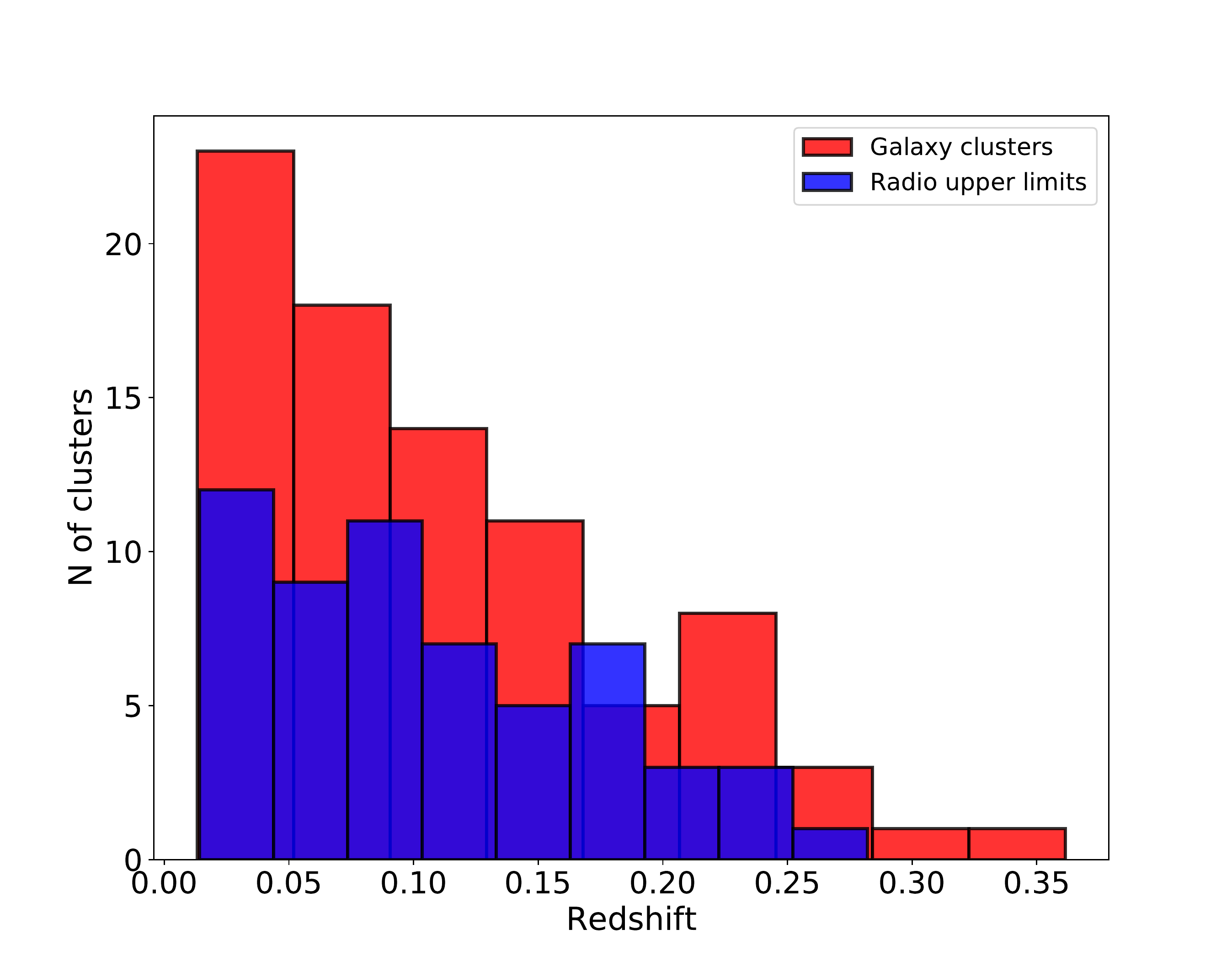}
	%\hspace*{0.2cm}
	\includegraphics[height=27em, width=28em]{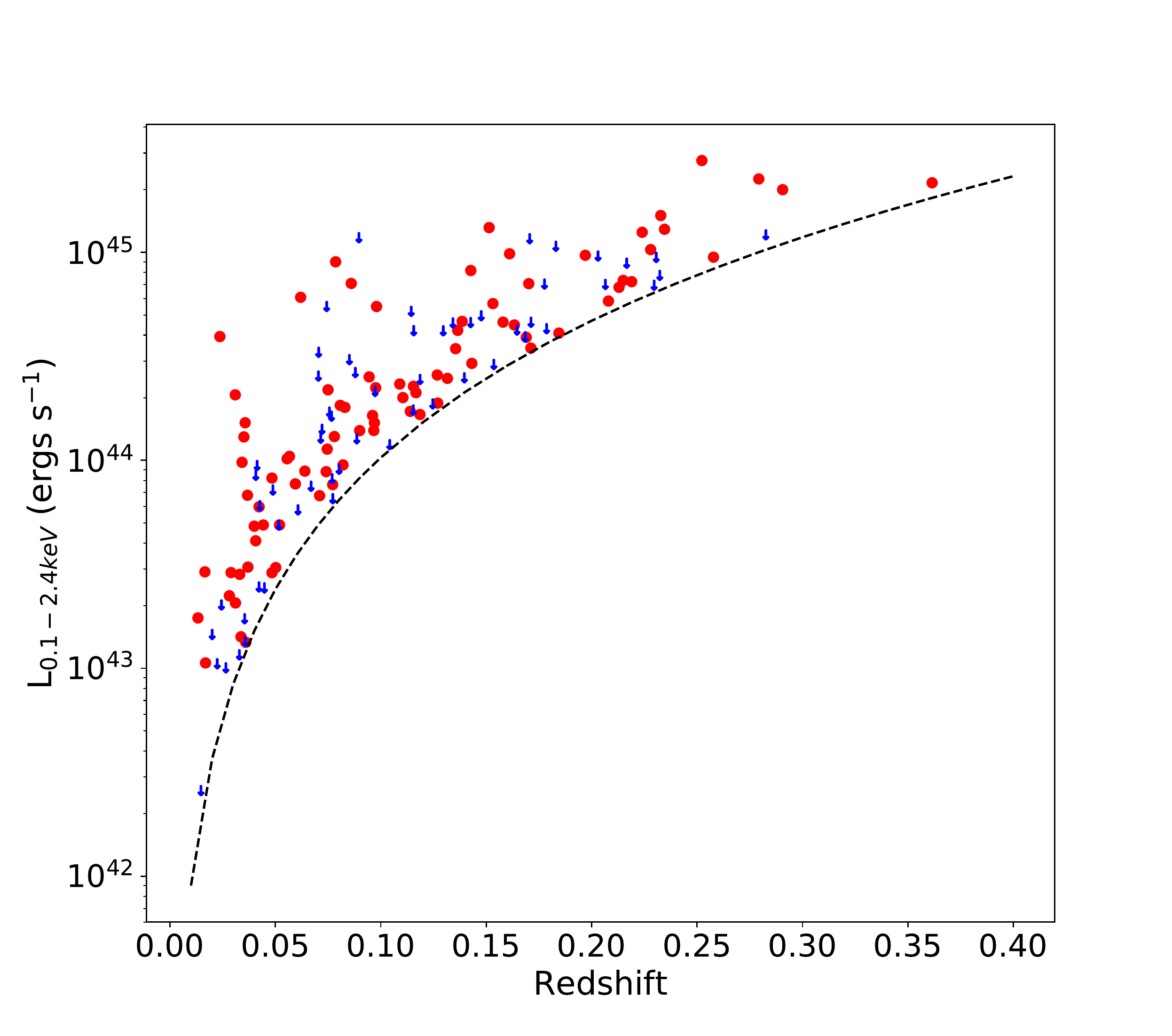}
	\caption{\textit{Top Panel}: Histogram showing the redshift distribution of the cluster sample. \textit{Bottom Panel}: X-ray luminosity vs. redshift for the cluster sample. The dashed line represents the theoretical flux cut.}
	\label{fig:plclusters}
\end{figure}

The top panel of Fig. \ref{fig:plclusters} shows the redshift distribution for the cluster sample. The redshift range is much narrower than the group sample, with most clusters lying at $z \leq$ 0.15 and only a few objects reaching $z \sim$ 0.35. On the other hand, the bottom panel shows the Malmquist bias. The sensitivity of the X-ray observations of the cluster sample is uniform for all the objects, leading to a smoother distribution with respect to the group sample.

\begin{figure}
	%\hspace*{-6.95cm} 
	\includegraphics[height=27em, width=28em]{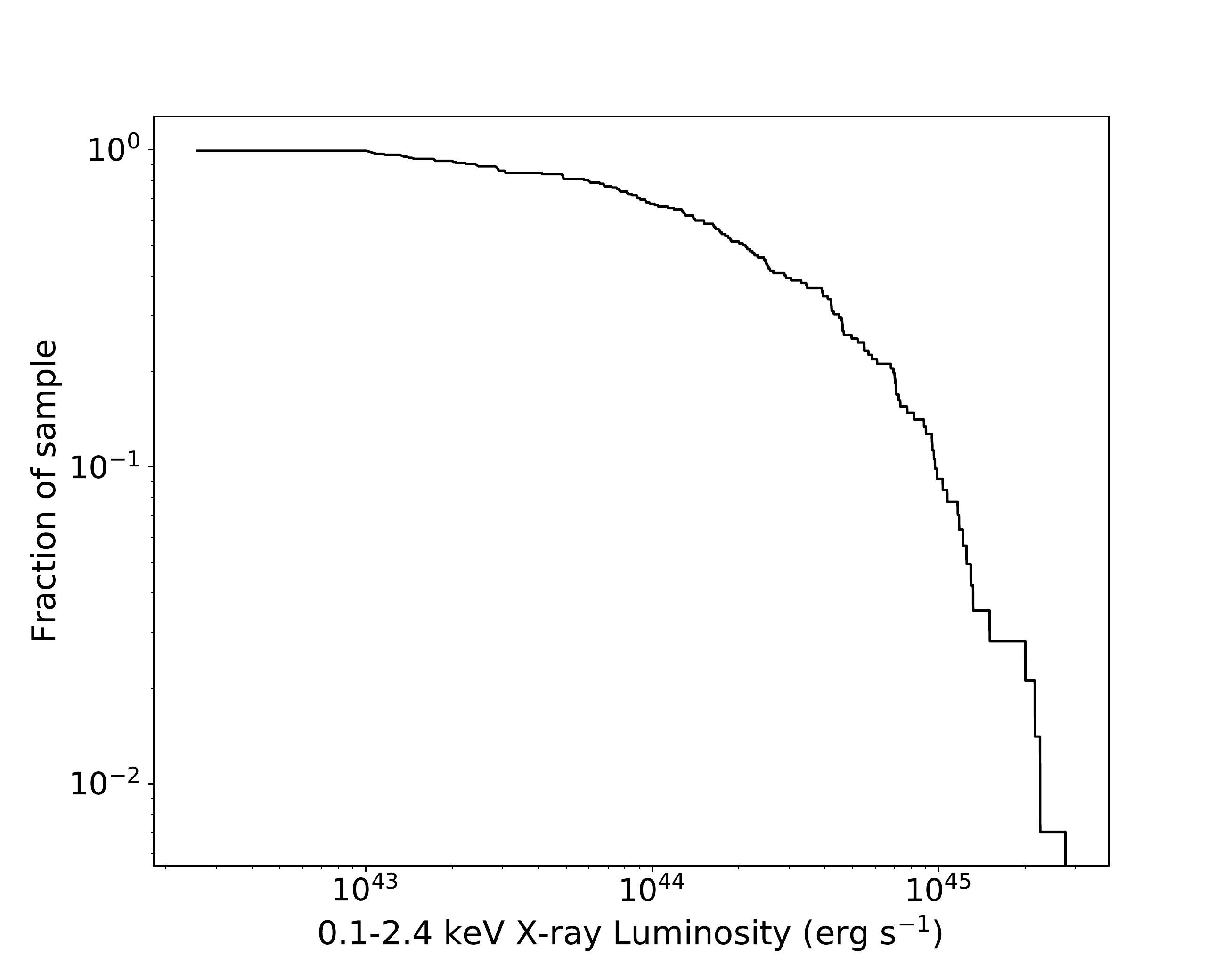}
	%\hspace*{0.2cm}
	\includegraphics[height=27em, width=28em]{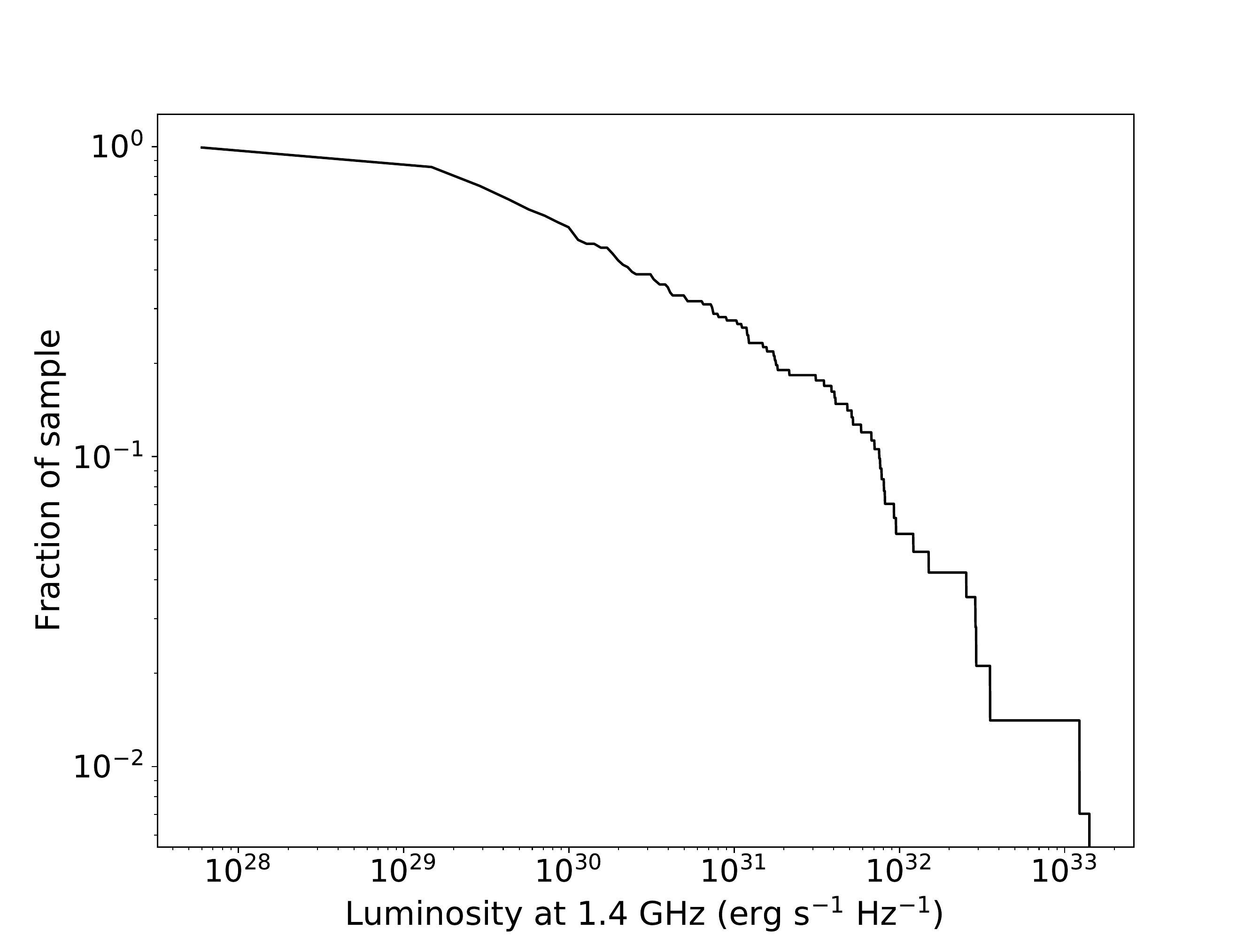}
	\caption{\textit{Top Panel}: X-ray luminosity distribution function for the cluster sample in the 0.1-2.4 keV band. \textit{Bottom Panel}: 1.4 GHz luminosity distribution function for the cluster sample.}
	\label{fig:funcclusters}
\end{figure}

Fig.~\ref{fig:funcclusters} shows the X-ray and radio luminosity distribution for the cluster sample. While the X-ray luminosity of groups never exceeds $\sim$ 10$^{44}$ erg s$^{-1}$, clusters are able to reach $\sim$ 5 $\cdot $10$^{45}$ erg s$^{-1}$. On the other hand, the 1.4 GHz power of clusters stops at  $\sim$ 10$^{33}$ erg s$^{-1}$ Hz$^{-1}$, while radio sources hosted in groups can, in some cases, be 10 times more powerful. We will discuss this in more depth in the following sections.

%%%%%%%%%%%%%%%%%%%%%%%%%%%%%%%%%%%%%%%%%%%%%%%%%%%%%%%%%%%%%%%%%%%

\section{Analysis} 
\label{sec:analysis}

In the following, we will focus on the relationship between log(P$_{1.4 GHz}$) vs log(L$_X$) for groups and clusters, where P$_{1.4 GHz}$ is the 1.4 GHz power of the central radio source and L$_X$ is the X-ray luminosity of the intra-group/cluster medium.

\subsection{Correlation between X-ray and radio luminosity}
\label{sec:stattests}

\begin{figure}
	%\hspace*{-6.9cm} 
	\centering
	\includegraphics[height=28em, width=29em]{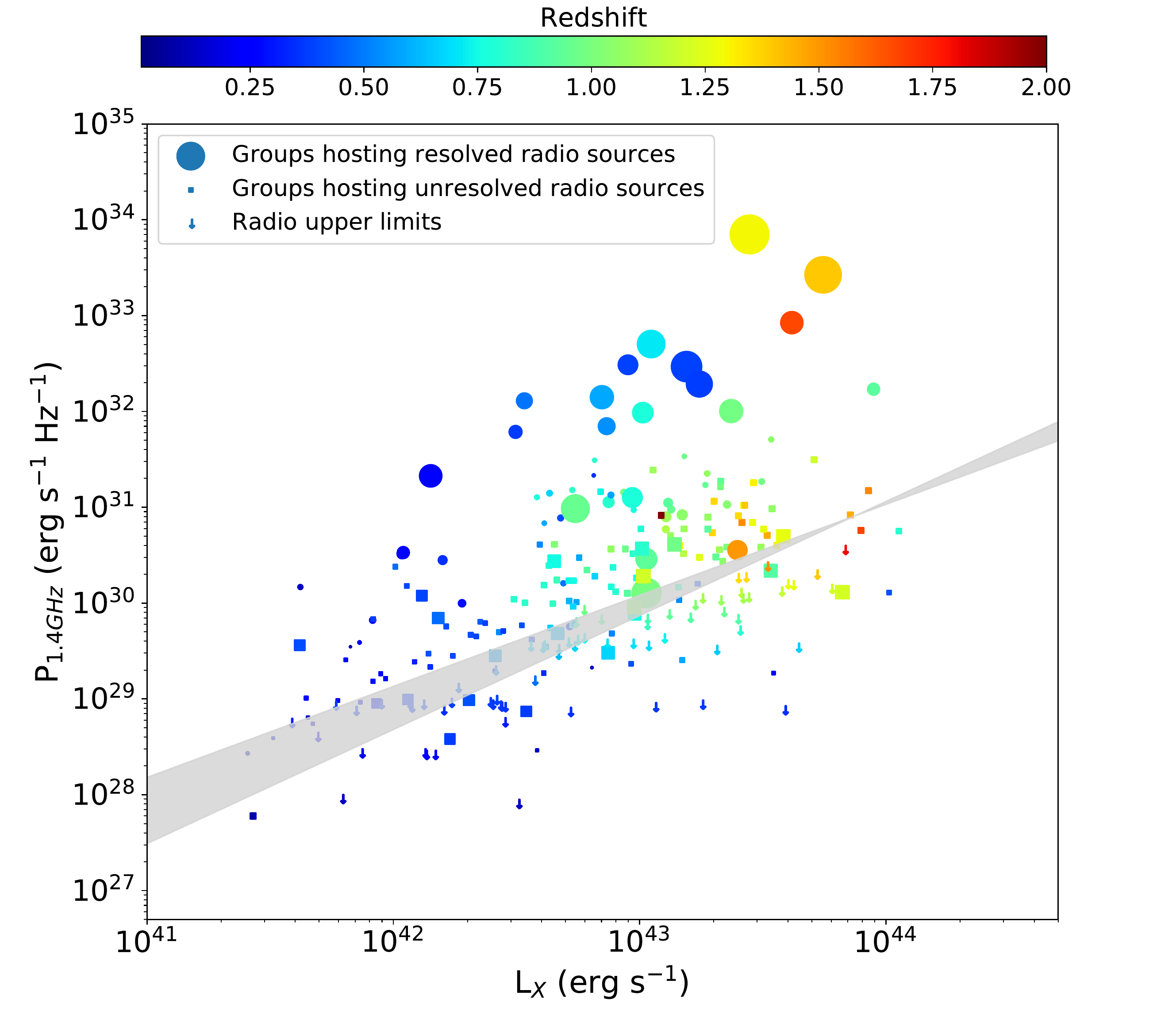}
	\caption{1.4 GHz luminosity of the central radio galaxy vs. intra-group medium X-ray luminosity in the 0.1-2.4 keV band for the group sample. The points are sized by the radio LLS and colorized for the redshift. Down arrows denote MIGHTEE radio upper limits. The grey area represents the best fit: $\log L_{\text{R}}$ = (1.07 $\pm \ 0.12) \ \cdot \log L_{\text{X}}$ - (15.90 $\pm \ 5.13$).}
	\label{fig:corr}
\end{figure}

In Fig.~\ref{fig:corr} we show the 1.4 GHz luminosity of the radio galaxy vs. the X-ray luminosity from the intra-group medium for every group, where the sizes of symbols are proportional to the LLS and colour denotes redshift. The upper limits in the radio band are represented by down-sided arrows.

\begin{figure}
	%\hspace*{-6.9cm} 
	\centering
	\includegraphics[height=28em, width=29em]{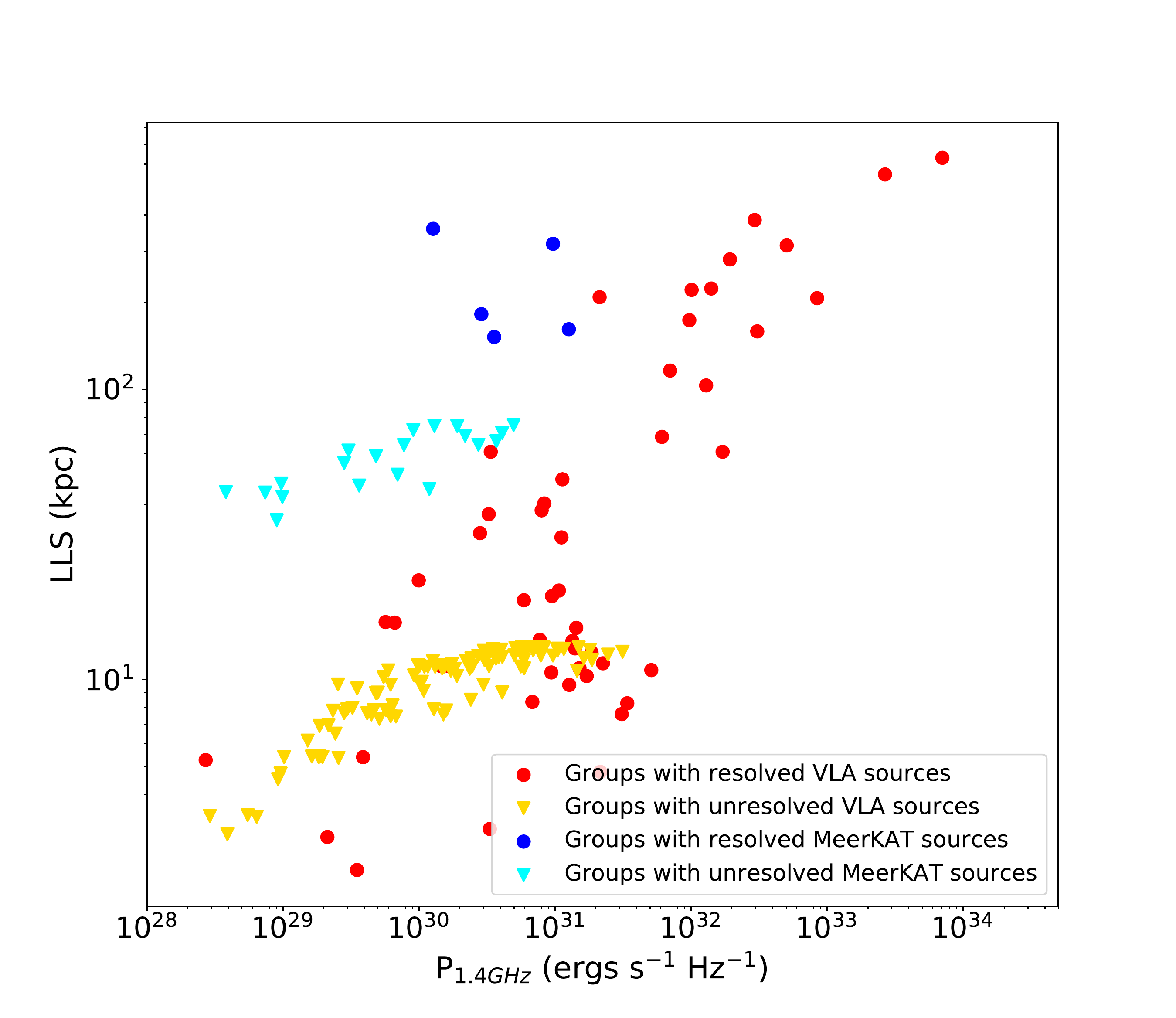}
	\caption{Largest Linear Size (LLS) vs. radio power at 1.4 GHz of the central radio galaxies for the group sample. Circles represent groups hosting resolved sources, while triangles are unresolved ones. For the latter class, the upper limits on the LLS were estimated as 1.5$\theta$, with $\theta$ being the resolutions of the VLA ($\sim$ 2.5\arcsec) and MeerKAT ($\sim$ 9\arcsec) observations.}
	\label{fig:lls}
\end{figure}

Groups hosting unresolved radio galaxies follow a narrow distribution (see squares in Fig. \ref{fig:corr}) that suggests a possible connection between radio and X-ray luminosities, with higher intra-group medium emission corresponding to higher power coming from the central radio source. However, groups hosting big radio galaxies (up to $\sim$ 600 kpc, see also Fig.~\ref{fig:lls}) with higher radio luminosities broaden the distribution. This is consistent with what we have previously argued from the radio distribution function (Fig.~\ref{fig:lumfunct}), in which the function for resolved objects is able to reach $\sim$ 10$^{34}$ erg s$^{-1}$ Hz$^{-1}$, while unresolved radio sources never go above $\sim$ 10$^{31}$ erg s$^{-1}$ Hz$^{-1}$.

As discussed above, the Malmquist bias is able to produce spurious correlations when two luminosities in different bands are compared. In order to test for Malmquist bias, we used the partial correlation Kendall $\tau$ test \citep{Akritas_Siebert_1996}, as in \citet{Ineson_2015}. This allows us to look for correlations between radio and X-ray luminosities in the presence of upper limits and a dependence on redshift. The results, with the null hypothesis being that the correlation is produced by the dependence on redshift, are presented in Table \ref{tab:kendall}.

\begin{table}%[!htb]
	%\smallsize
	\centering 
	\begin{tabular}{c c c c}
		\hline
		\hline
		Sample & N & $\tau/\sigma$ & $p$ \\
		\hline
        Galaxy groups & 247 & 5.04 & $<$ 0.0001 \\
        Galaxy groups, no res. groups & 197 & 3.94 & $<$ 0.0001 \\
        Galaxy groups, uniform flux cut & 175 & 4.16 & $<$ 0.0001 \\
        Galaxy clusters & 142 & 2.99 & 0.0028 \\
        Full sample & 389 & 6.66 & $<$ 0.0001 \\
        Full sample, no res. groups & 339 & 8.84 & $<$ 0.0001 \\
        \hline
	\end{tabular}
	\caption{Results of the partial correlation Kendall's $\tau$ test in the presence of a correlation with a third factor. N is the sample size, $\tau$ is the partial correlation statistic, $\sigma$ is the standard deviation, while $p$ represents the probability under the null hypothesis that the correlation is produced by the dependence on redshift.}
	\label{tab:kendall}
\end{table}

We found a strong correlation for the group sample, estimating $p  \ <$ 0.0001, with $p$ being the null hypothesis probability. However, as discussed in Sec. \ref{sec:stattests}, the X-ray minimum sensitivity used to build the catalog is not constant across the entire COSMOS field. Especially at low fluxes, there is the chance that this could lead to ambiguous results. To address this issue, we applied a further, uniform flux cut at 5 $\cdot$ 10$^{-15}$ erg s$^{-1}$ cm$^{-2}$, and repeated the test for this subsample. We still find $p  \ <$ 0.0001, supporting the hypothesis that there is a intrinsic correlation between X-ray and radio luminosities. We also performed the test excluding groups hosting resolved radio sources. In fact, as discussed in more detail in Sec.~\ref{sec:grg}, such objects could be characterized by a different balance between the X-ray and the radio emission, thus widening the correlation. We estimated, even for this subsample, $p  \ <$ 0.0001. %The same result was also found combining galaxy clusters and groups, while for clusters only we estimated $p  \ <$ 0.0028, that still suggests a strong correlation.
\\ \indent
However, \citet{Bianchi_2009} argued that the partial correlation Kendall's $\tau$ test may underestimate the redshift contribution to the relation, particularly when it comes to determining the functional correlation. On the other hand, they also showed that random scrambling of their radio luminosities is not able to produce the observed slope of the correlation with the X-ray luminosity, thus suggesting that a physical correlation may be present. We will return to this issue in Sec.~\ref{sec:simulation}.

\subsection{Comparison with the cluster sample}

\begin{figure}
	%\hspace*{-6.9cm} 
	\centering
	\includegraphics[height=28.5em, width=29.5em]{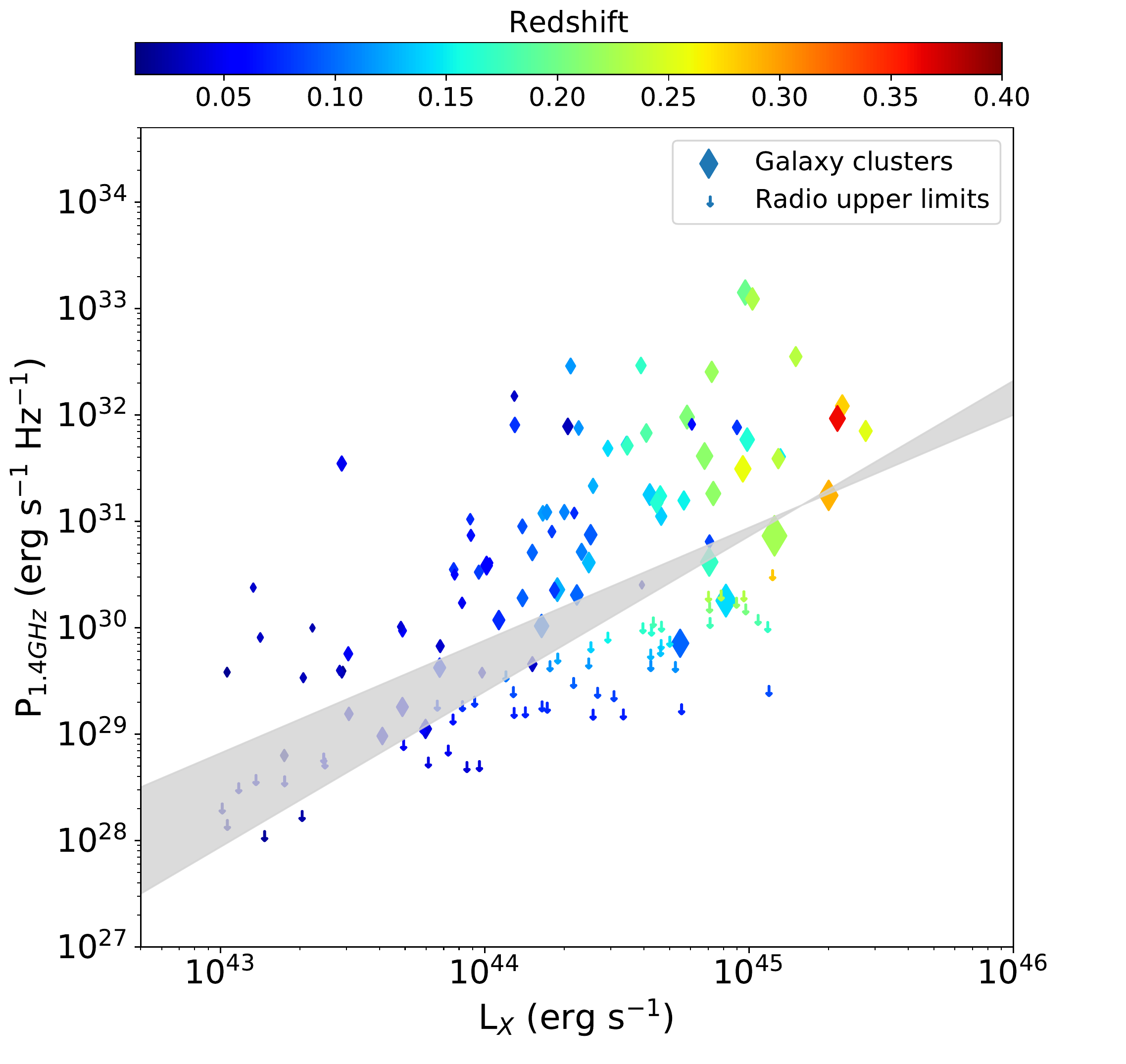}
	\caption{1.4 GHz luminosity of the central radio galaxy vs. ICM X-ray luminosity in the 0.1-2.4 keV band for the cluster sample. The points are sized by the radio LLS and colorized for the redshift. Down arrows denote NVSS radio upper limits. The grey area represents the best fit: $\log L_{\text{R}}$ = (1.26 $\pm \ 0.20$) $\cdot \  \log L_{\text{X}}$ - (25.80 $\pm \ 9.04$).}
	\label{fig:corrclusters}
\end{figure}

In Fig.~\ref{fig:corrclusters} we show the 0.1-2.4 keV ICM luminosity vs. the 1.4 GHz power of the central radio galaxy for the cluster sample. As in Fig.~\ref{fig:corr}, the sizes of symbols are proportional to the LLS, and colour denotes redshift. It appears that a correlation also exists in galaxy clusters, albeit with more scatter compared to groups hosting unresolved radio sources. However, unlike for groups, the LLS seems to be less correlated with the central radio power, and there is only one radio source with LLS $>$ 200 kpc. Since the cluster sample is restricted to lower redshifts, one could argue that large and powerful radio galaxies could be only found at $z \geq 0.3$. 
However, large radio sources in the group sample are already found at $z$ $\leq$ 0.3, where sources in the cluster sample are instead small (see Sec. \ref{sec:discussion} for a further discussion). Moreover, \citet{Gupta_2020} have shown that there is little redshift evolution in the radio luminosity at fixed host stellar mass out to $z\sim 1$ for sources hosted in clusters. However, their results do show strong mass evolution in the number of radio-powerful AGN.

Again we applied the Kendall $\tau$ test in order to check whether the X-ray/radio correlation is significant, with the results listed in Table \ref{tab:kendall}. We found a null-hypothesis probability of $p = 0.0028$. This suggests that  clusters also show a correlation, even though it is weaker than for groups. Finally, combining the two samples (group sample and cluster sample), we obtain $p  \ <$ 0.0001.

\subsection{The correlation for clusters and groups: comparison with simulated datasets}
\label{sec:simulation}
\begin{figure}
	%\hspace*{-6.9cm} 
	\centering
	\includegraphics[height=28.5em, width=29.5em]{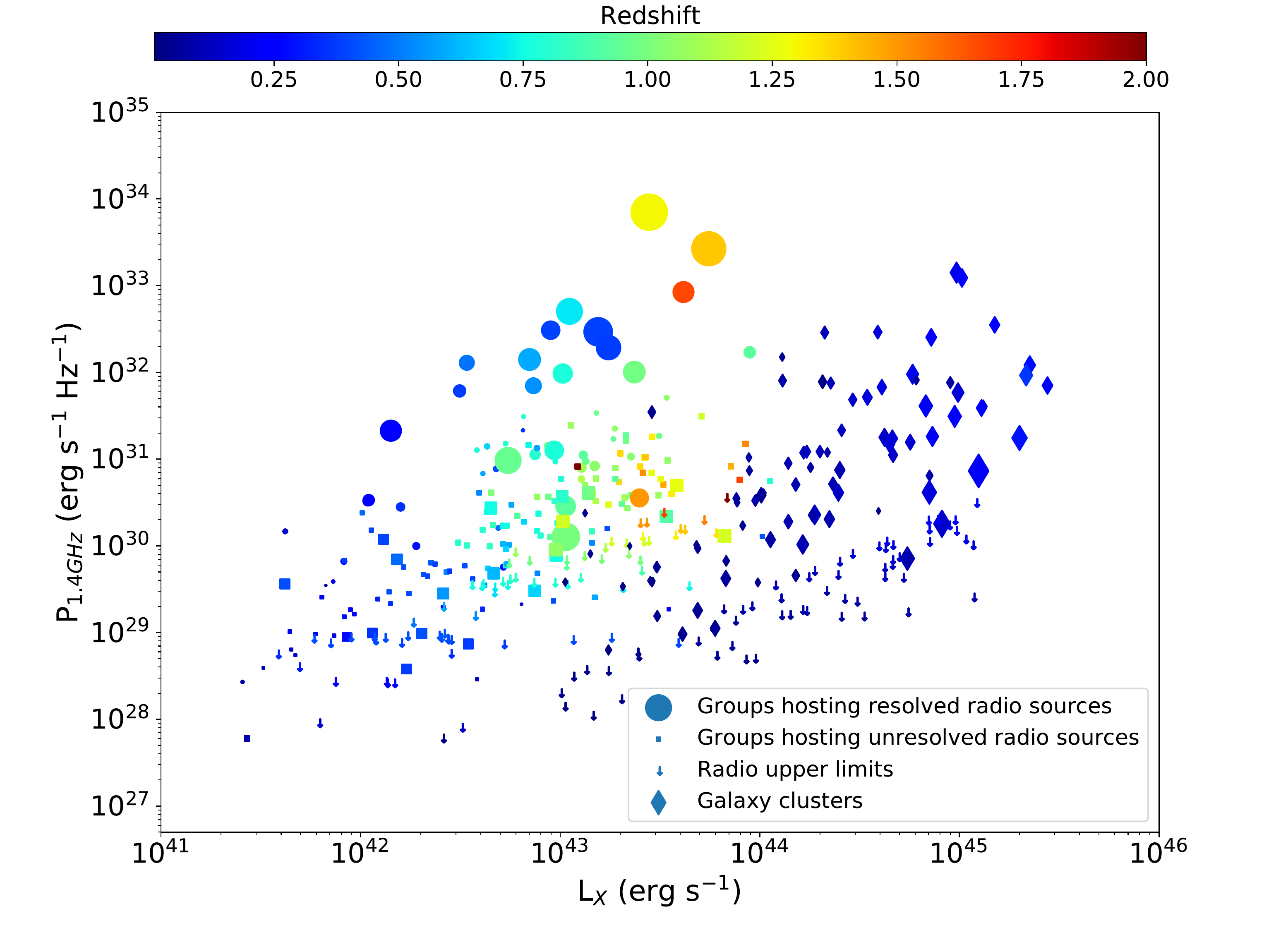}
	\caption{1.4 GHz luminosity of the central radio galaxy vs. ICM X-ray luminosity in the 0.1-2.4 keV band for both the group sample (circles) and the cluster sample (diamonds). The points are sized by the radio LLS and colorized for the redshift.}
	\label{fig:all}
\end{figure}

In Fig.~\ref{fig:all} we show the X-ray luminosities vs the 1.4 GHz powers for both the group sample and the cluster sample. Obvious is the dearth of data at L$_{\text{X}} \ < 5 \ \cdot$ 10$^{43}$ erg s$^{-1}$ - L$_{\text{1.4GHz}} \ \sim 10^{32}$ erg s$^{-1}$ Hz$^{-1}$, set by the flux cuts of our samples. Since we are dealing with left-censored data (i.e., upper limits), any correlation that depends on data where the lowest values are upper limits are hard to assess. This is a long-standing problem in astronomy \citep[see e.g.,][]{Feigelson_1992}.

\begin{figure}
	%\hspace*{-6.9cm} 
	\centering
	\includegraphics[height=28em, width=29em]{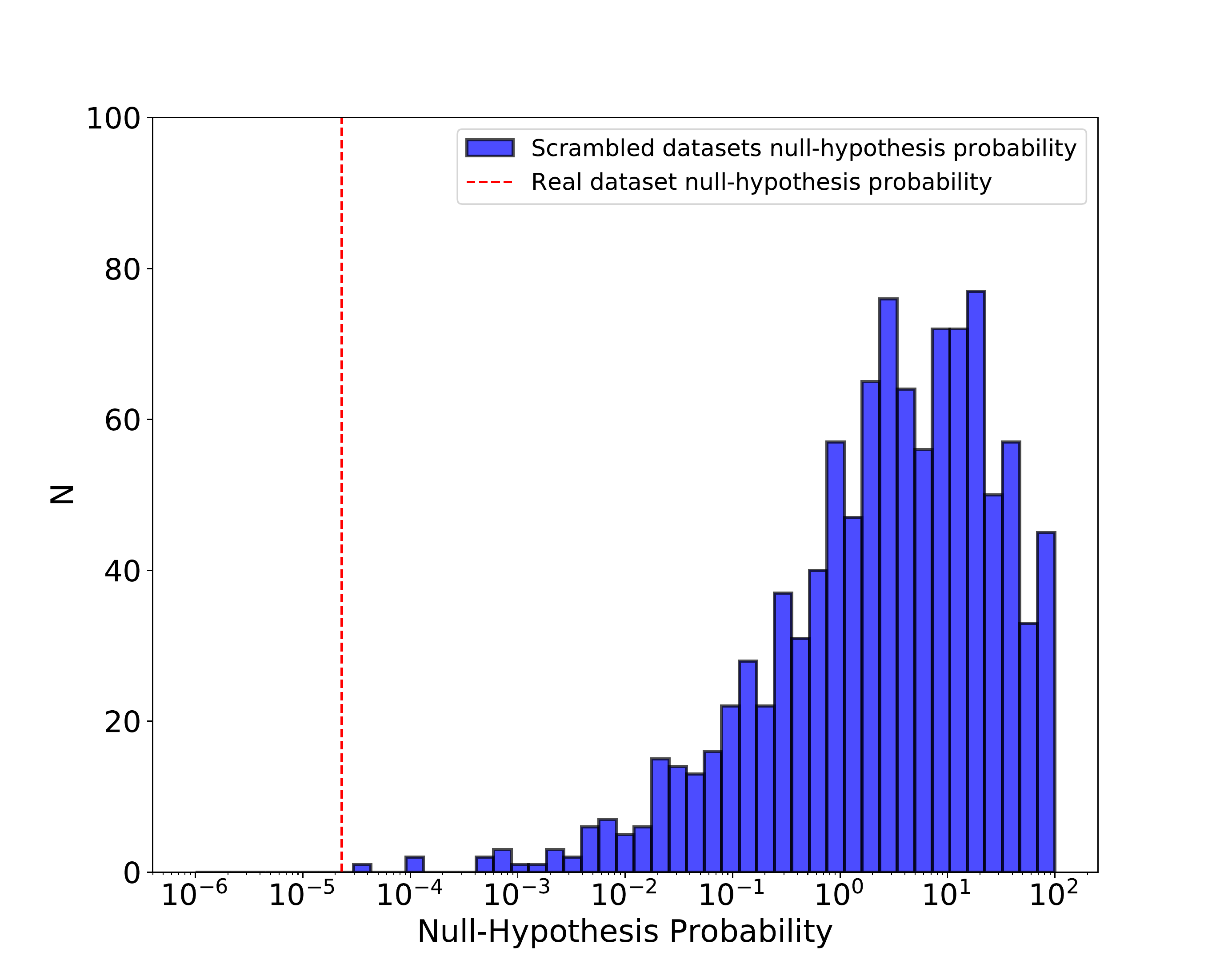}
	\caption{Null-hypothesis probability distribution of N=1000 scrambled datasets. The red dashed line represents the null-hypothesis probability of the real data.}
	\label{fig:pvalues}
\end{figure}

In order to investigate the effects of this left-censoring we performed a ``scrambling test'', that was applied to similar problems by, e.g., \citet{Bregman_2005} and \citet{Merloni_2006}. In this test we keep each pair of X-ray luminosity and redshift of the group sample and `shuffle' the associated radio fluxes, assigning each one to a random ($L_X$/$z$) pair. Applying the newly assigned redshift, we then calculate the radio luminosity from the flux. 
We produced 1000 scrambled datasets in this manner and calculated for each of them the null-hypothesis probability through the Kendall $\tau$ test. The distribution of such values is presented in Fig.~\ref{fig:pvalues}.

Out of 1000 'shuffling', the null-hypothesis probability was never found to be lower than the real dataset. The distribution is similar to a lognormal, with the peak lying between $\sim$ 2\% and 20\%. The mean sets around $\sim$ 12.5 \%, with a standard deviation of $\sim$ 3.5 \%. The null-hypothesis probability of the real dataset lies more than 3$\sigma$ away from it, suggesting that the correlation holds.

%%%%%%%%%%%

Alternatively, we can also simulate our population of radio and X-ray sources by randomly drawing them from luminosity functions and redshift distributions, then applying the respective flux cuts in the X-ray and radio and measuring the correlation. This Monte-Carlo simulation can be performed with and without an underlying correlation between X-ray and radio powers.

The simulation was performed along the following steps:

\begin{itemize}

    \item We first draw a random redshift within the ranges $z=0.01-2$ for the group sample and $z=0.01-0.4$ for the cluster sample, assuming a constant comoving source density, i.e.:
    
    \begin{equation}
    \hspace{2.5cm}
        \dfrac{dN}{dz}=\dfrac{4\pi c {D_L}^2(z)}{(1+z)H_0 E(z)} ,
    \end{equation}
    \\
    
    with $N$ being the number of objects, $c$ the speed of light, $E$(z)= $\sqrt{\Omega_M(1+z)^3+ %\Omega_K(1+z)^2+ 
    \Omega_\Lambda}$ and ${D_L}$ the luminosity distance at redshift $z$. \\

    \item We then sampled $N$ X-ray luminosities, assuming two different luminosity functions as probability density functions (pdf). For clusters, we assumed the $0.1-2.4$ keV luminosity function (LF) of BCS \citep{Ebeling_1998b}, since the cluster sample was built starting from this catalog. On the other hand, for groups we used the LF presented in \citet{Koens_2013} that goes down to lower luminosities and out to $z \sim$ 1.1. Multiple studies have shown hints of negative redshift evolution of the LF, with a reduction in the number density of massive, luminous clusters at high redshifts \citep[e.g.,][]{Moretti_2004, Koens_2013}. As usual, they use a Schechter function in the form:

   \begin{equation}
   \hspace{2.5cm}
   \Phi(L) = \Phi^* \exp(-L/L^*) \, L^{-\alpha}   ,
   \end{equation}
   
where $L$ is the X-ray luminosity in units of 10$^{44}$ erg s$^{-1}$, $\Phi^*$ is the normalization, $L^*$ is the luminosity at the function cutoff and $\alpha$ determines the steepness of the function at L $<$ L$^*$. Following the same approach described in \citet{Koens_2013} and \citet{Bohringer_2014}, the redshift evolution is taken into account by parametrizing density and luminosity evolution through a power-law:

\begin{equation}
\hspace{2.5cm}
 \Phi^*(z) = \Phi_0^* (1+z)^A   ,
\end{equation}

\begin{equation}
\hspace{2.5cm}
 L^*(z) = L_0^* (1+z)^B   ,
\end{equation}

where $\Phi_0^*$ and $L_0^*$ are the values at the current epoch, A $\sim$ -1.2 and B $\sim$ -2 \citep{Moretti_2004}. As we first draw a redshift, we can then sample the X-ray luminosity. The number of mock objects, $N$, was chosen to match the numbers in our samples.\\
    
    \item We estimated the flux and applied an X-ray flux cut of 2 $\cdot \ 10^{-15}$ erg s$^{-1}$ cm$^{-2}$ for groups and 4.4 $\cdot \ 10^{-12}$ erg s$^{-1}$ cm$^{-2}$ for clusters, as for our samples. \\
    
    \item We associated every X-ray luminosity-redshift pair with a radio luminosity assuming (i) no correlation between X-ray and radio power, (ii) $\log L_{\text{R}}$ = (1.07 $\pm \ 0.12) \ \cdot \log L_{\text{X}}$ - (15.90 $\pm \ 5.13$) for both clusters and groups, and (iii) the above correlation for groups, and $\log L_{\text{R}}$ = (1.26 $\pm \ 0.20$) $\cdot \  \log L_{\text{X}}$ - (25.80 $\pm \ 9.04$) for clusters. The correlations were estimated exploiting the parametric EM algorithm coded in the AStronomical SURVival statistics package (ASURV, \citealt{Feigelson_2014}), that takes into account different contributions by detections and left-censored data.
    
\end{itemize}

\begin{figure}
	\centering
	\includegraphics[height=23em, width=28.5em]{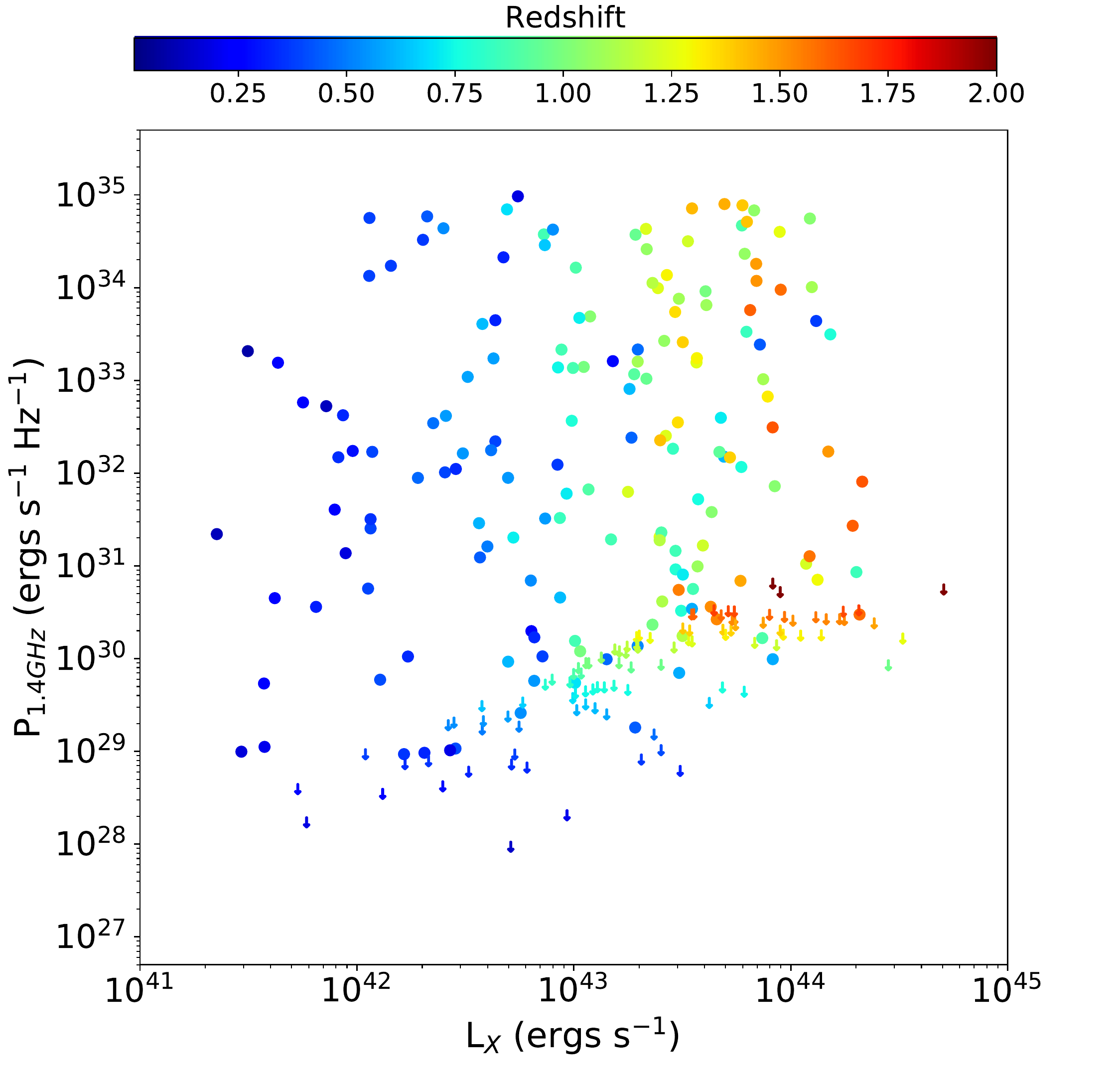}
	\includegraphics[height=24em, width=28.5em]{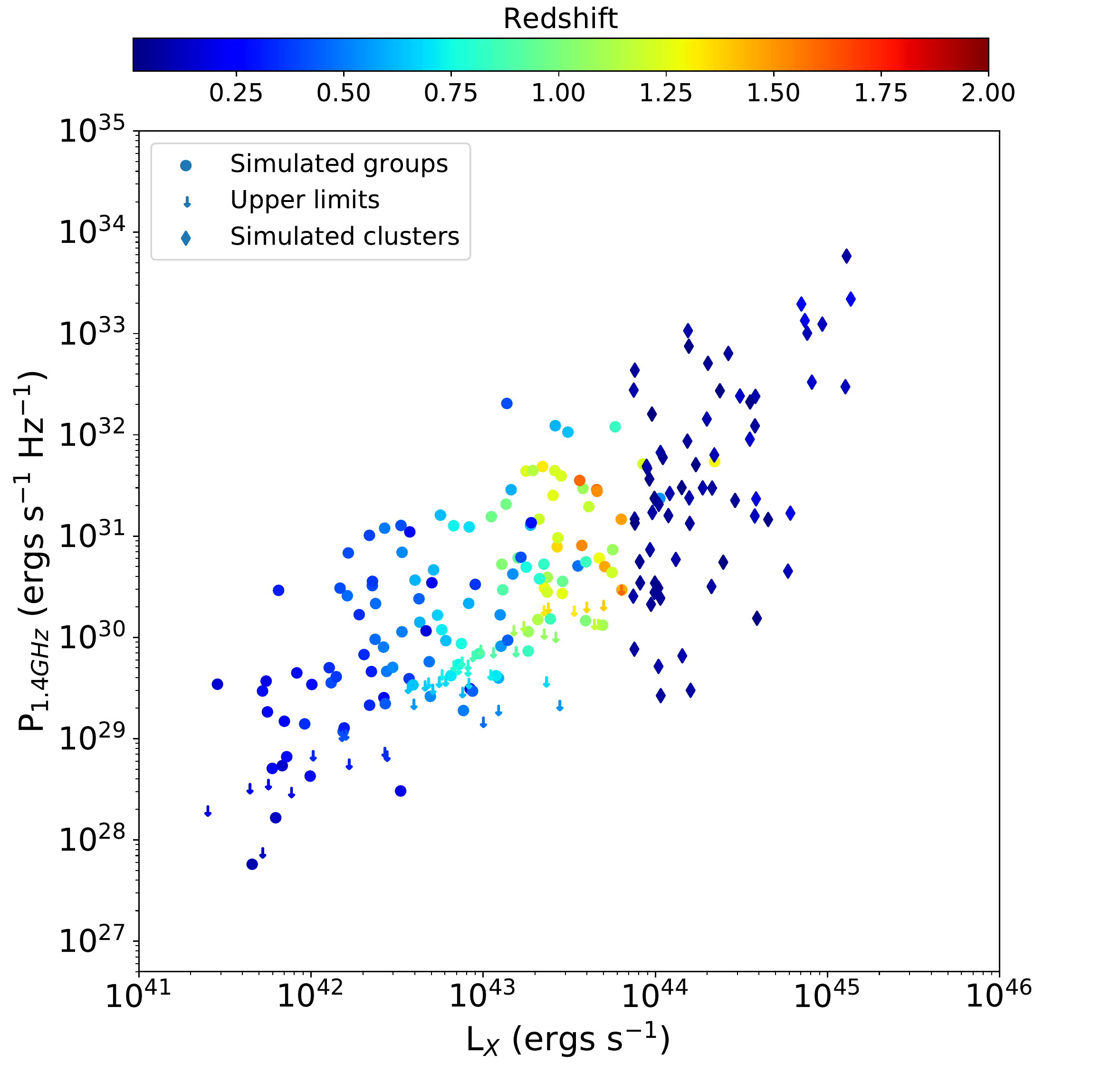}
	\includegraphics[height=24em, width=28.5em]{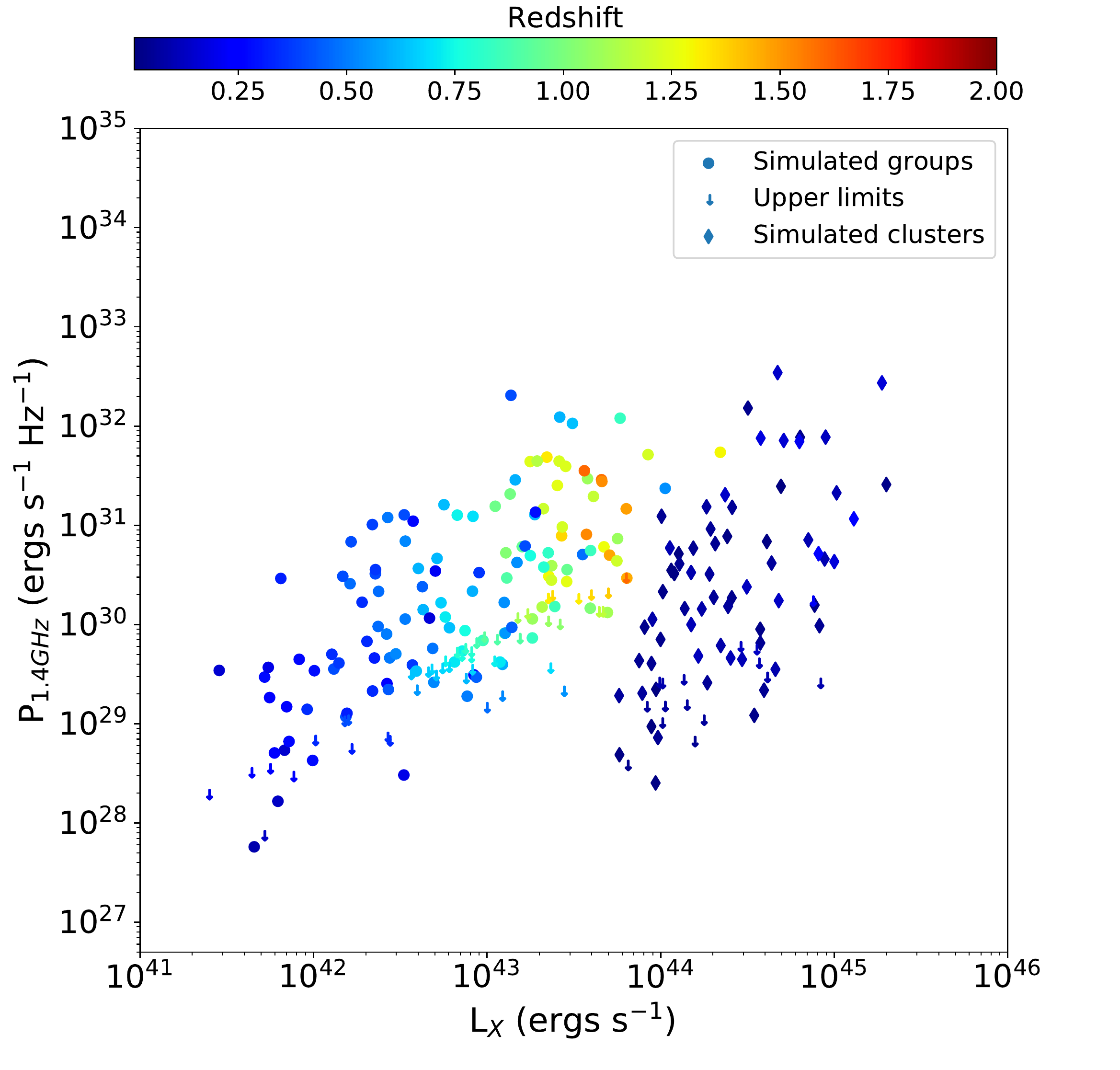}
	\caption{1.4 GHz power of the central radio source vs. X-ray luminosity of the ICM in the 0.1-2.4 keV band for the simulated data. \textit{Top panel}: No correlation. \textit{Middle panel}: Same correlation (see text) for both galaxy clusters (diamonds) and groups (circles). \textit{Bottom panel}: Two different correlations for galaxy clusters (diamonds) and groups (circles).}
	\label{fig:simulation}
\end{figure}

The results for groups assuming no correlation between X-ray and radio power are presented in the top panel of Fig.~\ref{fig:simulation}. The empty bottom-right area of the plot is produced by the Malmquist bias. Apart from this, the distribution of simulated data looks significantly different from both the group sample and the cluster sample. In fact, Fig.~\ref{fig:corr} shows no groups with L$_{\text{X}} <$ 10$^{43}$ erg s$^{-1}$ hosting a radio galaxy with L$_{\text{R}}  >$ 10$^{33}$ erg s$^{-1}$ Hz$^{-1}$. However, we can see such objects in the top panel of Fig.~\ref{fig:simulation}, despite the simulation having the same flux cuts of the real observation. This means that groups are prevented by physical limitations from being found at these luminosities: the lack of them in this region is not produced by biases.

%\begin{figure}
%	%\hspace*{-6.9cm} 
%	\centering
%	\includegraphics[height=26em, width=29em]{onecorr.pdf}
%	\caption{X-ray luminosity of the ICM in the 0.1-2.4 keV band vs. 1.4 GHz power of the central radio source for the simulated data, applying the same correlation for both galaxy clusters (diamonds) and groups (circles). The relation, estimated through a linear regression on the group sample, is $\log L_{\text{R}}$ = (1.07 $\pm \ 0.12) \ \cdot \log L_{\text{X}}$ - (15.90 $\pm \ 5.13$).}
%	\label{fig:onecorr}
%\end{figure}

The middle panel of Fig.~\ref{fig:simulation} shows simulated data assuming the same correlation for clusters and groups. The plot still looks different from Fig.~\ref{fig:all}, in which low X-ray luminosity clusters are offset from groups, producing a tail that stands out from the distribution.
In order to properly measure the differences between the simulated and the real distributions, we used the two sample Kolmogorov-Smirnov test. The test estimates the null-hypothesis probability that two samples belong to the same distribution. The X-ray luminosities of the simulated clusters are drawn from the BCS luminosity function, that is the same catalog used for building the cluster sample. Since the KS test is not accurate when performed on two samples if one of them is drawn from the other one, we chose to perform it only on radio luminosities. Comparing the simulated data of the middle panel of Fig.~\ref{fig:simulation} with the real dataset, the KS test gives a null-hypothesis probability of $\sim$ 2.7\%, confirming that the two distributions are intrinsically different.
This suggests that X-ray and radio luminosities of clusters and groups could follow different correlations.

%\begin{figure}
%	%\hspace*{-6.9cm} 
%	\centering
%	\includegraphics[height=26em, width=29em]{diffcorr.pdf}
%	\caption{X-ray luminosity of the ICM in the 0.1-2.4 keV band vs. 1.4 GHz power of the central radio source for the simulated data, applying two different correlations for galaxy clusters (diamonds) and groups (circles). The relation for clusters, obtained through a linear regression on the cluster sample, is $\log L_{\text{R}}$ = (1.26 $\pm \ 0.20$) $\cdot \  \log L_{\text{X}}$ - (25.80 $\pm \ 9.04$), while for groups we assumed $\log L_{\text{R}}$ = (1.07 $\pm \ 0.12) \ \cdot \log L_{\text{X}}$ - (15.90 $\pm \ 5.13$).}
%	\label{fig:diffcorr}
%\end{figure}

Finally, the bottom panel of Fig.~\ref{fig:simulation} shows the results of the simulation assuming two different correlations for clusters and groups. The distribution of the simulated data is now similar to the real correlation, suggesting that this is the assumption that better fits our data. This is also confirmed by the KS test, that gives a null-hypothesis probability of $\sim$ 39.1\%, indicating that the two samples likely belong to the same distribution.
We also argue that, without providing any relation between radio and X-ray luminosity, the simulation is not able to reproduce the correlation, indicating that the redshift is not the only factor contributing the correlation. This supports the picture of a physical connection between ICM and AGN emission.
%galaxy groups follow a narrow distribution, suggesting the existence of a tighter correlation when compared to the one of clusters, that is broader. 

%%%%%%%%%%%%%%%%%%%%%%%%%%%%%%%%%%%%%%%%%%%%%%%%%%%%%%%%%%%%%%%%%%%%%%%%%%%%%%%%%%%%%%%%%%%%

\section{Discussion}
\label{sec:discussion}

\subsection{Large radio galaxies in galaxy groups}
\label{sec:grg}

Fig.~\ref{fig:corr} and \ref{fig:lls} show that AGN in the centre of groups typically reach only tens of kpc in linear size and up to $\sim$ 10$^{32}$ erg s$^{-1}$ Hz$^{-1}$ in power. However, a few of them ($\sim$ 10) are able to grow to hundreds of kpc, reaching radio powers comparable to massive clusters' BCGs (Fig.~\ref{fig:corrclusters}) and surpassing such sources in terms of size. We then argue that the biggest radio galaxies are found in the centre of galaxy groups. This was already hinted by multiple works on giant radio galaxies (e.g., \citealt{Mack_1998, Machalski_2004, Subrahmanyan_2008, Chen_2012, Grossova_2019, Cantwell_2020}, and references therein). However, this is the first time that they are included as sources in a large sample of groups, and their link to the environment is studied with respect to $"$classical$"$ radio sources. Large radio galaxies have been found in galaxy clusters, too. An example is the giant radio fossil recently observed in the Ophiucus clusters \citep{Giacintucci_2020} that reaches a size of $\sim$ 1 Mpc. However, they look more like outliers produced by unusually energetic AGN outburst, rather than widespread cases. This is also supported by Fig. \ref{fig:all}, which shows that the LLS of radio galaxies hosted in clusters is usually significantly smaller with respect to sources found in groups. The cluster sample only shows one cluster with a central radio source bigger than 200 kpc ($\sim$ 0.7\% of the sample), while the group sample has 10 of them ($\sim$ 4\% of the sample), ranging from 200 to 600 kpc.
\\ \indent
We suggest that the lower gas density in groups when compared to clusters and a different, more efficient accretion mechanism could be responsible for this effect. This is supported by \citet{Ineson_2013, Ineson_2015}, who found a very similar relationship between intra-group medium and AGN luminosities and showed that large radio galaxies, lying in the top region of Fig. \ref{fig:corr}, are mostly HERGs, powered by a radiatively efficient accretion mode \citep[e.g.,][]{Best_2012}, while groups lying within the narrower radio power distribution mostly contain LERGs. Similarly, we tried to classify the radio galaxies of the group sample into HERGs and LERGs, by retrieving optical spectra from the zCOSMOS survey \citep{Lilly_2005} and measuring the equivalent width (EW) of the [OIII] line. Objects showing EW$_{[\text{OIII}]} \ <$ 5 \AA \ were classified as LERGs, while radio galaxies with EW$_{[\text{OIII}]} \ >$ 5 \AA \ are HERGs. However, we were able to properly perform such analysis only on 9 radio galaxies, 4 of which were classified as HERGs, and 5 as LERGs. This was due to most objects not having a zCOSMOS detection, while for others the fit of the spectrum was inconclusive. Therefore, since this sample is too small to draw any conclusion from it, we will not discuss it and leave it for future follow-up works.

\begin{figure}
	%\hspace*{-6.9cm} 
	\centering
	\includegraphics[height=28em, width=29em]{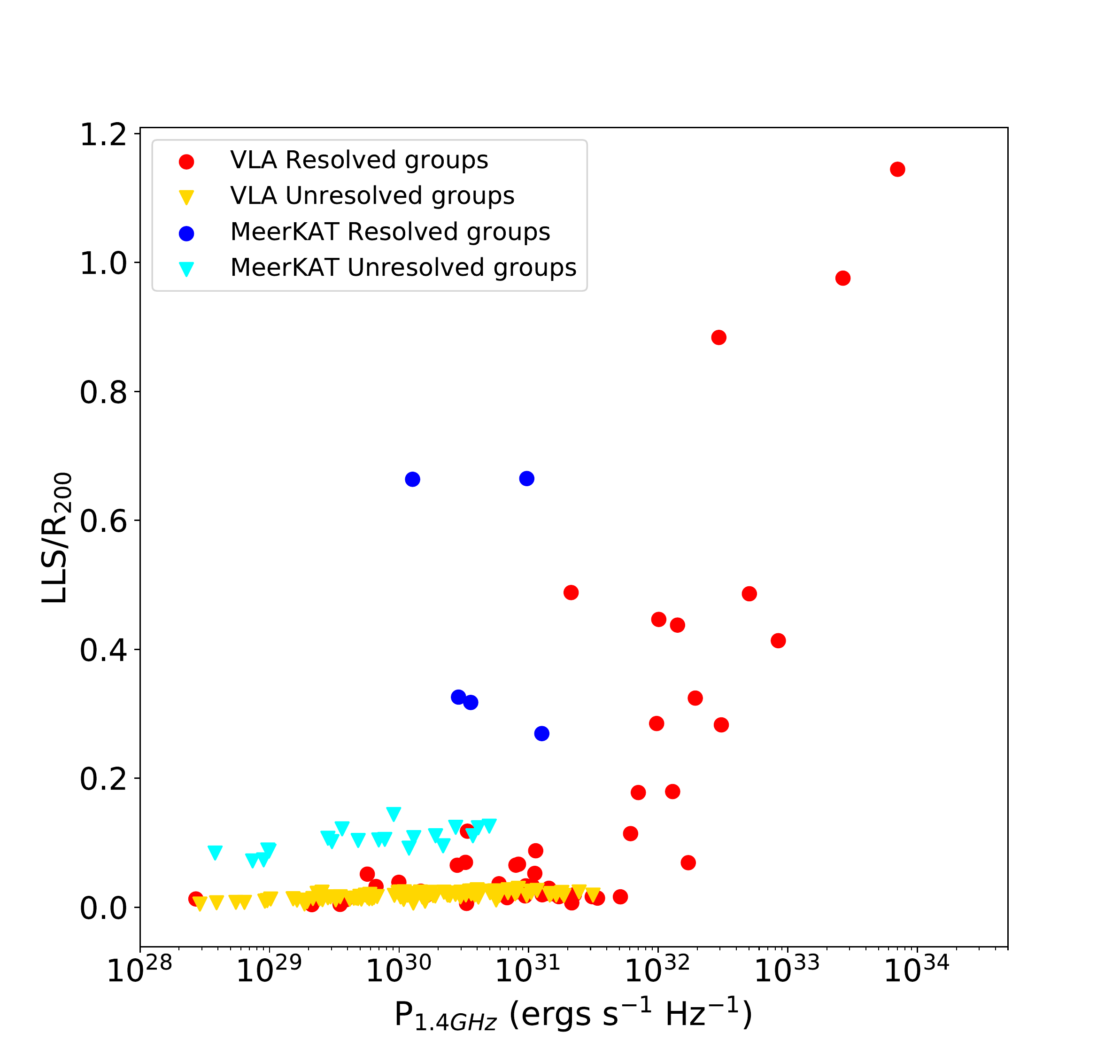}
	\caption{Ratio of LLS of the radio galaxy and R$_{200}$ of corresponding group (in kpc) vs. radio power of the radio galaxy. Circles represent groups hosting resolved radio sources, while triangles denote unresolved ones.}
	\label{fig:llsr200}
\end{figure}

Fig.~\ref{fig:llsr200} shows the radio power of the radio galaxy on the $x$-axis, and the LLS and corresponding group's R$_{200}$ ratio on the $y$-axis. The largest radio galaxies have dimensions comparable to the group's virial radius. 

%\begin{figure}
%	\hspace*{-0.5cm} 
%	\centering
%	\includegraphics[height=25em, width=28em]{rg.pdf}
%	\caption{The biggest radio galaxies have dimensions that are comparable to those of the group they are hosted in. An example is J095822.93+022619.8, with a linear size of 660 kpc, hosted in a group whose virial radius is 550 kpc. The radio image is from VLA-COSMOS Deep at 1.4 GHz; the beam is 2.5$"$$\times$2.5$"$, the sensitivity $\sim$ 12 $\mu$Jy beam$^{-1}$.}
%	\label{fig:rg}
%\end{figure}

The power output supplied by such AGN could potentially have strong consequences for the intra-group medium. However, the energy is transferred to the diffuse gas at large distances from the centre ($\sim$ hundreds of kpc), due to the dimensions of the radio galaxy, while the cores of  groups are typically only $\sim$ tens of kpc. This means that, when heating mainly occurs through radio mode feedback (i.e., by generating cold fronts and X-ray cavities), the centres of groups are less affected by such process, and thus cooling could be less suppressed than in galaxy clusters. On the other hand, when shocks are the dominant source of heating, it could still be enough to adequately quench radiative losses. This scenario looks very similar to some galaxy clusters, that are shown to host giant cavities, with lobes extending over the cooling region \citep[e.g.,][]{Gitti_2007a}. Some analyses of singular groups already suggested that jets extending well over the core could violate the standard AGN feeding-feedback model \citep[e.g.,][]{O'Sullivan_2011c, Grossova_2019}. However, this is the first work to prove that a significant fraction of the galaxy groups population effectively shows hints that support this scenario.
\\ \indent
This hypothesis needs to be tested by performing a thorough study of the jets and structure of these large radio sources, and possibly by comparing the results with a accurate analysis of the diffuse gas within the cooling radius of the host group. Such a radius is usually defined as the radius within which the cooling time of the ICM falls under the lookback time at $z$=1, corresponding to 7.7 Gyr. While meticolous estimates of the cooling radius are usually feasible for galaxy clusters, there are only a few, closeby groups \citep[e.g.,][]{O'Sullivan_2017} that have been observed with the required depth and resolution to accurately measure it. Therefore, we are currently unable to perform this comparison for our sample.
%This is also supported by deep observations of fossil galaxy groups (groups with absence of a recent group-scale merger), that show that the AGN energy injection is not able to properly balance the ICM radiative losses \citep[e.g.,][]{Jetha_2007, Miraghaei_2014}. 
%High angular resolution, X-ray studies of single objects are required to investigate deeper on this hypothesis.

%%%%%%%%%%%%%%%%%%%%%%%%%%%%%%%%%%%%%%%%%%%%%%%%%%%%%%%%%%%%%%%%%%%%%%%%%%%%%%%%%%%%%%%%%%%%
\subsection{Do clusters and groups show the same X-ray - radio correlation?}
\label{sec:cooltime}

In the previous sections we showed how, according to our analysis, the relation between the X-ray luminosity and the power of the central radio source is not produced by biases or selection effects. Clusters and groups seem to follow two different correlations, albeit with a similar slope. Furthermore, central radio galaxies in some groups show enhanced emission up to $\sim$ 3 orders of magnitude. The hypothesis of two, distinct correlations is supported by the analysis previously performed on our simulated datasets.

However, we suggest that clusters and groups could follow the same correlation, in which the latter populate the low X-ray and low radio luminosity regions, while the former are usually stronger. The correlation is mainly produced by groups and clusters that have not recently experienced a significant interaction with the surrounding environment, while those which have undergone recent mergers or accretion from other objects tend to broaden the distribution. Specifically, we would expect these clusters to show a lower AGN power at a given X-ray luminosity, since the lack of cooling ICM prevents the AGN from accreting gas. This could also explain the low-radio luminosity tail of clusters. The result of this scenario is a distribution that is narrower for galaxy groups, and then broadens because of the difference between cool cores and merging clusters. 
%However, it is not easily possible to include the effects of clusters dynamics in the simulation. Therefore, the code would need two different correlations to reproduce the observed data.

Recent results by \citet{Gupta_2020} (hereafter G20) already suggested the existence of a link between the large-scale properties of clusters and AGN feedback. Using SUMSS data (843 MHz) for two different cluster samples ($\sim$ 1000 X-ray selected, $\sim$ 12000 optically selected), they have shown that the probability of a cluster hosting radio-loud AGN scales with its mass. This could suggest a connection between AGN feedback and cluster mass. A similar link was already explored by \citet{Hogan_2015}. G20 also found \textit{no} evidence that the AGN radio power scales with the cluster halo mass (see Fig. 10 of G20). However, their sample is built taking into account every radio AGN within clusters, even those not hosted in BCGs. Therefore, it is not straightforward to compare it to the properties of central radio sources.

\begin{figure}
	%\hspace*{-6.9cm} 
	\centering
	\includegraphics[height=29em, width=29.5em]{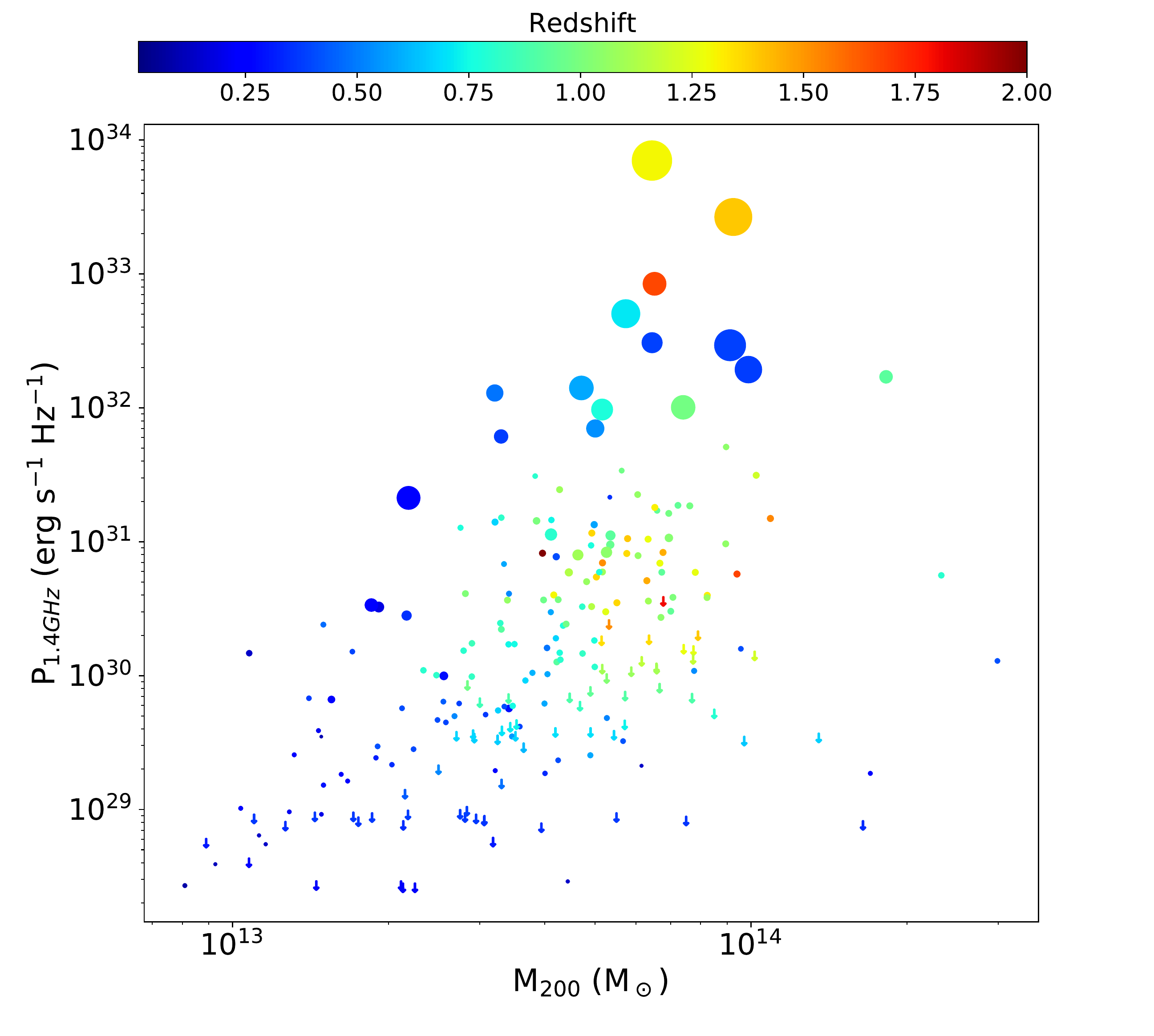}
	\caption{AGN power at 1.4 GHz for the group sample vs. mass (M$_{200}$). Colors denote redshift, and the points are sized for the LLS of the radio source. Down arrows represent radio upper limits.}
	\label{fig:massradio}
\end{figure}

In Fig.~\ref{fig:massradio} we show the AGN radio power at 1.4 GHz vs. the mass of the group sample (M$_{200}$). We find no hints of groups under M$_{200} \sim$ 6 $\cdot$ 10$^{13}$ M$_{\odot}$ hosting radio sources with log P (erg s$^{-1}$ Hz$^{-1}$) $>$ 32. This suggest that, even in the group regime, low-mass objects usually host weaker radio sources, while radio-loud AGN are usually found at higher masses.

One could argue that our sample is simply missing low-mass groups that host powerful AGN. These objects could therefore be rare. Since we do not detect them with a $\sim$ 250 objects sample, the probability of observing one has to be lower than 0.4\%. This suggests that, even if they do exist, they can be considered outliers. In the group sample, the only process we found that could bring to these consequences is the combination of low density medium and the more efficient accretion mechanism that produces large radio galaxies. (see Sec. \ref{sec:grg}). However, all these sources are found at higher masses. Either low-mass groups hosting radio-loud AGN do not exist or, if they do, they are extremely rare ($p \ <$ 0.4\%).

In order to support the hypothesis that clusters and groups follow the same correlation, we were able to classify 38 clusters of the cluster sample ($\sim$ 30\%) into cool-core and non-cool core. The classification, based on the presence of optical line emission, is presented by \citet{Hogan_2015}.

\begin{figure}
	%\hspace*{-6.9cm} 
	\centering
	\includegraphics[height=29em, width=28.5em]{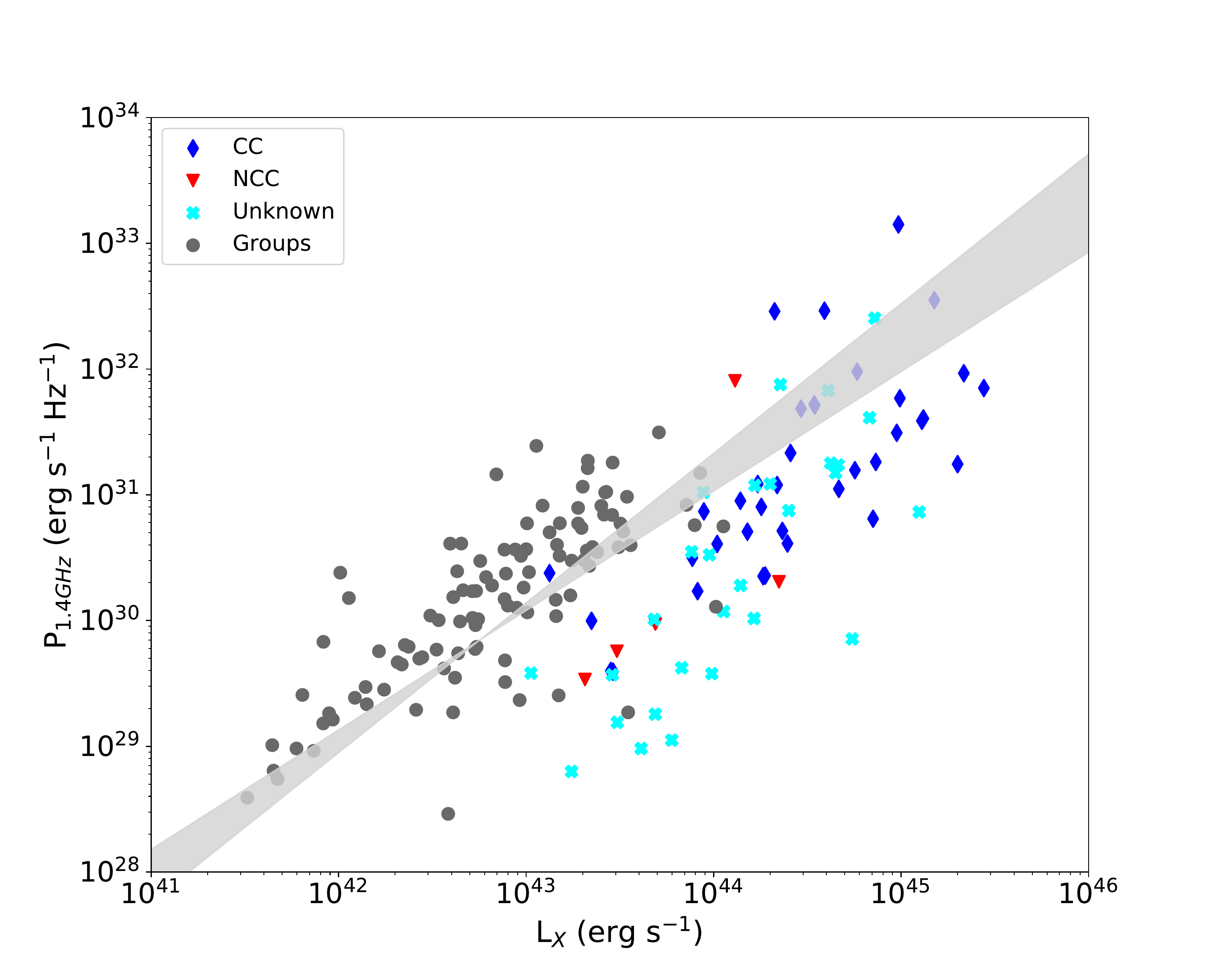}
	\caption{1.4 GHz power of the central radio source vs. X-ray luminosity in the 0.1-2.4 keV band for 71 of the cluster sample objects, classified into cool cores (blue diamonds), non-cool cores (red triangles) and clusters with unknown dynamical state (cyan crosses). Grey circles represents groups hosting unresolved radio sources, for comparison. The grey area is the best-fit relation for galaxy groups: logL$_{\text{R}}$ = (1.07 $\pm \ 0.12) \ \cdot \log \text{L}_{\text{X}}$ - (15.90 $\pm \ 5.13$).}
	\label{fig:coolcores}
\end{figure}

There are 33 clusters for which we have no information about their dynamical state. Not surprisingly, most cool cores populate the high-luminosity area of the plot. Furthermore, 4 out of 5 non-cool cores are found at low radio power. We thus argue that cool cores produce the high-power tail of the correlation shown in Fig.~\ref{fig:all}, while more dynamically disturbed clusters tend to broaden it. 

%However, a relationship between the total X-ray luminosity and the current power of the central radio galaxy is much harder to address, since we are comparing a large and long-time scale property (such as the total ICM emission of the cluster or group) with a short-time scale process, that can thus significantly vary during the same amount of time. Nevertheless, a correlation similar to our case was already pointed out in other works; an example is \citet{Ineson_2015}, that found a link between the X-ray luminosity and the 150 MHz power of a sample of radio galaxies within clusters and groups; \citet{Hogan_2015} discussed the same, broad trend for their sample of BCGs, arguing that more massive objects likely tend to host more powerful AGNs and, in turn, more massive SMBHs.
%If the correlation really exists, we expect it to be wide, since every object can undergo different evolutionary processes that can now favour the X-ray emission, now power the AGN; and since the time scales are so different, it takes time for one process to have consequences on the other.

One possible scenario is that the feedback cycle in galaxy groups is much tighter and the central AGN can affect the entropy of the gas more efficiently, as seen in our group sample. Over their lifetime, central radio galaxies can undergo phases of higher power, allowing them to grow to hundreds of kpc. Once central radio galaxies have grown to a certain size, the energy from the AGN is injected at greater radii with respect to the cooling radius (see Sec.~\ref{sec:grg}), weakening the efficiency of feedback and broadening the correlation between X-ray luminosity and radio power. Via mergers and accretion, groups can grow to become clusters \citep{White-Frenk_1991}. Events occurring during their lifetime, such as mergers, accretion or any interaction with other objects, that significantly affect both the cooling of the ICM and the central radio galaxy, can widen the distribution.

\section{Conclusions}
\label{sec:conclusions}

We have studied the correlation between the X-ray emission of the intra-group medium and the radio power of the central AGN for a sample of 247 X-ray selected galaxy groups detected in the COSMOS field. We compared the properties of these groups with a control sample of galaxy clusters and with simulated datasets of X-ray and radio luminosities. Our conclusions can be summarized as follows:

\begin{itemize}
    \item Groups show a correlation between the intra-group medium emission and the radio galaxy power, with more X-ray luminous objects hosting more powerful radio sources. Using Kendall's partial correlation $\tau$ test, combined with data 'scrambling' and Monte-Carlo simulations,  we showed that this correlation is not produced by the flux cut. Groups hosting large radio sources ($\geq$ 150-200 kpc) stand out of the correlation.\\
    \item Galaxy clusters are usually more luminous but show a similar correlation to groups, albeit with more scatter. \\
    \item Despite the observational evidence of two, distinct correlations, we argue that clusters and groups could follow the same correlation once the dynamical state is taken into account. Groups populate the low-luminosity region, while cool-core clusters are found at high luminosities.\\
    \item Mergers between galaxy clusters and the resulting changes to central cooling times, as well as changes in the accretion mechanism, can increase the scatter in the observed correlation.\\
    \item Galaxy groups host a significantly higher fraction of large (LLS $>$ 200 kpc) radio galaxies ($\sim$ 4\%) than clusters ($\sim$ 0.7\%), albeit the redshift range of our cluster sample is narrower. The growth of these radio sources, that in this work reach up to $\sim$ 600 kpc, is probably favored by the low ambient densities and aided by an efficient accretion mode (HERGs). Radiative cooling of the diffuse thermal gas could be less suppressed in these objects, since the AGN energy injection happens at larger radii compared to the cooling region of the corresponding group.\\
\end{itemize}

More detailed studies are needed to address these results. Volume-limited catalogs are essential in order to reduce biases that can be introduced by redshift-dependent luminosity functions and other effects. Deep, high angular resolution observations of single groups and optical spectra could also be helpful for a better understanding of the physical link between the intra-group medium and the central AGN.

\section*{Acknowledgements}\label{acknowledgments}
The authors thank the anonymous referee for useful comments and suggestions. T. Pasini is supported by the DLR Verbundforschung under grant number 50OR1906. We acknowledge discussions with J. Liske and J. Mohr. The spectral analysis on zCOSMOS data was performed by M. Owers. The MeerKAT telescope is operated by the South African Radio Astronomy Observatory, which is a facility of the National Research Foundation, an agency of the Department of Science and Innovation. We acknowledge use of the Inter-University Institute for Data Intensive Astronomy (IDIA) data intensive research cloud for data processing. IDIA is a South African university partnership involving the University of Cape Town, the University of Pretoria and the University of the Western Cape.

\section*{Data Availability}
The data underlying this article are available in the article and in its online supplementary material.

\newcommand{\newblock}{}
\bibliographystyle{aasjournal}
\bibliography{bibliography}

\begin{appendices}
\section{Properties of the group sample}
\label{appendix}

The properties of the group sample are listed in Table \ref{tab:complete}. The full table is available as online material, while a part of it is showed here; the listed properties include X-ray coordinates of the group, redshift, M$_{200}$, R$_{200}$, X-ray luminosity, %name of the central radio source in VLA-COSMOS Deep (when detected with VLA), 
radio coordinates, flux density at 1.4 GHz, radio luminosity at 1.4 GHz and LLS (when resolved). The last column reports 1 if the radio source was detected with both VLA and MeerKAT, and 2 if it was detected only by MeerKAT.

\begin{sidewaystable*}
    \footnotesize
    \vspace{6cm}
	\begin{tabular}{c c c c c c c c c c c c}
		\hline
		\hline
		RAJ2000$_X^{(1)}$ & DECJ2000$_X^{(2}$ & z$^{(3)}$ & M$_{200}^{(4)}$ & R$_{200}^{(5)}$ & L$_X^{(6)}$ & RAJ2000$_R^{(7)}$ & DECJ2000$_R^{(8)}$ & S$_{1.4}^{(9)}$ & L$_{1.4}^{(10)}$ & LLS$^{(11)}$ & Detection$^{(12)}$ \\
		
		[deg] & [deg] & & [10$^{13}$ M$_\odot$] & [kpc] & [10$^{42}$ erg s$^{-1}$] & & & [mJy] & [10$^{30}$ erg s$^{-1}$ Hz$^{-1}$] & [kpc] \\
		\hline
        \hline
        150.5111 & 2.02699 & 0.899 & 3.86 $\pm$ 1.10 & 513.45 & 8.624 $\pm$ 4.052 & 10:02:02.549 & +02:01:45.36 & 0.441 $\pm$ 0.040 & 14.31 $\pm$ 1.56 & 15.06 & 1 \\
        150.62251 & 2.16039 & 1.5 & 6.52 $\pm$ 1.70 & 500.37 & 41.48 $\pm$ 18.12 &  10:02:30.117 & +02:09:12.45 & 7.751 $\pm$ 0.010 & 844.3 $\pm$ 0.01 & 206.85 & 1 \\
        150.57957 & 2.47898 & 0.61 & 3.21 $\pm$ 0.72 & 539.28 & 4.306 $\pm$ 1.597 & 10:02:18.308 & +02:28:04.29 & 1.081 $\pm$ 0.043 & 14.02 $\pm$ 0.64 & 12.79 & 1  \\
        150.17097 & 2.52363 & 0.697 & 2.75 $\pm$ 0.59 & 495.76 & 3.826 $\pm$ 1.359 & 10:00:41.418 & +02:31:24.17 & 0.716 $\pm$ 0.028 & 12.71 $\pm$ 0.58 & 9.57 & 1 \\
        149.83842 & 2.67517 & 0.26 & 2.56 $\pm$ 0.54 & 569.03 & 1.901 $\pm$ 0.665 & 09:59:21.341 & +02:40:30.45 & 0.546 $\pm$ 0.082 & 0.995 $\pm$ 0.16 & 21.96 & 1 \\
        ... & & & & & & & & & & & \\
        \hline
	\end{tabular}
    \caption{The table lists all the properties of the group sample and of their central radio sources, when detected. The first 5 lines of it are shown here; the columns of the table are: X-ray RA (1) and DEC (2) of the group, redshift (3), M$_{200}$ (4) with error in 10$^{13}$ M$_\odot$, R$_{200}$ (5) in kpc, X-ray luminosity (6) and error in 10$^{42}$ erg s$^{-1}$ in the 0.1-2.4 keV band, radio RA (7) and DEC (8), flux density at 1.4 GHz in mJy (9) with error, radio luminosity at 1.4 GHz (10) in 10$^{30}$ erg s$^{-1}$ Hz$^{-1}$ with error, and LLS in kpc (when resolved) (11). The last column (12) reports 1 if the radio source was detected with both VLA and MeerKAT, and 2 if it was detected only by MeerKAT.}
    \label{tab:complete}
\end{sidewaystable*}

\end{appendices}

\bsp	% typesetting comment
\label{lastpage}
\end{document}